\documentclass[apj,revtex4]{emulateapj}
\usepackage{graphicx, bm,placeins, natbib}
\usepackage{lscape}
\providecommand{\e}[1]{\ensuremath{\times 10^{#1}}}

\begin{document}

\title{X-ray Detected Active Galactic Nuclei in Dwarf Galaxies at $0<\MakeLowercase{z}<1$}
\author{K. Pardo\altaffilmark{1, $\star$}, A. D. Goulding\altaffilmark{1}, J. E. Greene\altaffilmark{1}, R.S. Somerville\altaffilmark{2}, E. Gallo\altaffilmark{3}, R.C. Hickox\altaffilmark{4}, B.P. Miller\altaffilmark{5}, A.E. Reines\altaffilmark{6,$\dagger$}, \& J.D. Silverman\altaffilmark{7}}
\affil{$^1$ Department of Astrophysical Sciences, Princeton University, 4 Ivy Lane, Princeton, NJ 08544, USA; kpardo@astro.princeton.edu}
\affil{$^2$ Department of Physics and Astronomy, Rutgers University, 136 Frelinghuysen Road, Piscataway, NJ 08854, USA}
\affil{$^3$ Department of Astronomy, University of Michigan, 1085 South University Avenue, Ann Arbor, MI 48109, USA; Hubble Fellow}
\affil{$^4$ Department of Physics and Astronomy, Dartmouth College, 6127 Wilder Laboratory, Hanover, NH 03755, USA}
\affil{$^5$ Department of Chemistry and Physical Sciences, The College of St. Scholastica, Duluth, MN 55811, USA}
\affil{$^6$ National Optical Astronomy Observatory, 950 North Cherry Avenue, Tucson, AZ 85719, USA}
\affil{$^7$ Kavli Institute for the Physics and Mathematics of the Universe (WPI), The University of Tokyo Institutes for Advanced Study, The University of Tokyo, Kashiwa, Chiba 277-8583, Japan}
\altaffiltext{$\star$}{kpardo@astro.princeton.edu}
\altaffiltext{$\dagger$}{Hubble Fellow}

\begin{abstract}
We present a sample of accreting supermassive black holes (SMBHs) in dwarf galaxies at $z<1$. We identify dwarf galaxies in the NEWFIRM Medium Band Survey with stellar masses $M_{\star}<3\e{9} M_{\odot}$ that have spectroscopic redshifts from the DEEP2 survey and lie within the region covered by deep (flux limit of $\sim 5\times 10^{-17} - 6\times 10^{-16} \rm{erg \ cm}^{-2} \ \rm{s}^{-1}$) archival \textit{Chandra} X-ray data. From our sample of $605$ dwarf galaxies, $10$ exhibit X-ray emission consistent with that arising from AGN activity. If black hole mass scales roughly with stellar mass, then we expect that these AGN are powered by SMBHs with masses of $\sim 10^5-10^6 \ M_{\odot}$ and typical Eddington ratios $\sim 5\%$. Furthermore, we find an AGN fraction consistent with extrapolations of other searches of $\sim 0.6-3\%$ for $10^9 \ M_{\odot} \leq M_{\star} \leq 3\e{9} \ M_{\odot}$ and $0.1<z<0.6$. Our AGN fraction is in good agreement with a semi-analytic model, suggesting that as we search larger volumes we may use comparisons between observed AGN fractions and models to understand seeding mechanisms in the early universe.
\end{abstract}


\section{Introduction}
Supermassive black holes (SMBHs; median mass of $10^8 \ M_{\odot}$) are ubiquitous in massive, bulge-dominated galaxies \citep[e.g.][]{Kauffmann2003}, and the masses of these black holes have been found to correlate with the properties of the stellar spheroid \citep[e.g., the galaxies' stellar velocity dispersion, $M_{\rm{BH}} - \sigma_{\star}$; e.g.,][]{McConnell2012, Kormendy2013}. These relationships suggest that a full understanding of galaxy evolution requires knowledge of SMBH formation and evolution. Unfortunately, there is very little known about SMBH formation and detecting the first SMBHs directly is practically impossible with the tools currently available. 

Low-mass SMBHs ($M_{\rm{BH}}\leq 10^6 \ M_{\odot}$) in dwarf galaxies ($M_{\star} \leq 3\times 10^9 \ M_{\odot}$, or approximately the mass of the Large Magellanic Cloud), which are expected to have had little growth through mergers or accretion, can provide an indirect window onto the primordial seeds of the SMBHs in the more massive galaxies we see today \citep{Bellovary2011}. Therefore, the distribution of central black hole masses in dwarf galaxies can give us many insights into SMBH and galaxy evolution \citep[see reviews in][]{Volonteri2010, Greene2012}. 

It is unclear whether all dwarf galaxies harbor massive black holes. From dynamical measurements, it is believed that M33 must have a supermassive black hole no larger than $1500 \ M_{\odot}$, if it has one at all \citep{Gebhardt2001}. Likewise, any black hole in NGC 205 must be less than $3.8\e{4} \ M_{\odot}$ \citep{Valluri2005}. On the other hand, there is dynamical evidence for SMBHs in some nearby dwarf galaxies, such as NGC404 and NGC 4395, which appear to host black holes with masses $\sim 10^5 \ M_{\odot}$ \citep[see, for example,][]{Filippenko1989,Seth2010, Thornton2008, denBrok2015}. In order to understand the connection between SMBHs and dwarf galaxies, we need a large, representative sample of dwarf galaxies with central SMBHs.

With current technology, direct dynamical measurements of $<10^5 \ M_{\odot}$ black holes cannot extend beyond a few Mpc. On the other hand, accreting SMBHs, known as active galactic nuclei (AGN), can be identified up to very large redshifts. The first attempts to search for a large sample of AGN in dwarf galaxies used broad and/or narrow optical emission lines to find AGN \citep[e.g.][]{Greene2004, Greene2007,Dong2012,Reines2013,Moran2014}. However, optical data is inherently limited by dust obscuration and emission from star formation \citep{Goulding2009}. Both mid-IR \citep{Satyapal2008, Satyapal2009, Sartori2015} and radio \citep{Reines2011,Reines2014a} searches have yielded complementary samples for follow-up \citep{Whalen2015,Reines2012}. In this paper, we focus on X-rays, which are insensitive to obscuration by dust and provide a relatively clean tool for identifying accreting black holes \citep[e.g.,][]{Brandt2015}. Searches of local dwarf galaxies in the X-ray have also yielded interesting targets \citep{Desroches2009,Schramm2013,Lemons2015}. \cite{Miller2015} used a uniform survey of early-type galaxies within $30$ Mpc to constrain the distribution of black hole masses in dwarf galaxies, but were fundamentally limited by their very small volume.

In order to find the best estimate of the occupation fraction of SMBHs in dwarf galaxies, larger survey volumes must be used, which implies we must push to higher redshifts. Recent deep surveys now allow us to complete a search for SMBHs in dwarf galaxies up to $z \sim 1$. X-ray stacking has been used to search for dwarf galaxies up to $z<1.5$ \citep{Mezcua2016}, but using the technique of looking for X-ray point sources that correspond to dwarf galaxies beyond $z\sim 0.4$ has not yet been attempted. Searching for X-ray point sources that coincide with dwarf galaxies at these higher redshifts requires deep X-ray data, as well as accurate galaxy masses in the same region. 

We have performed a search for SMBHs in dwarf galaxies up to redshift $z \lesssim 1$ in order to shed light on SMBH and galaxy evolution. In this paper, we outline our search for SMBHs in dwarf galaxies identified using a combination of NEWFIRM Medium-Band Survey (NMBS), \textit{Hubble Space Telescope} (\textit{HST}), \textit{Chandra} X-ray Observatory, \& DEEP2 observations. In Section 2, we outline our source selection process. We begin by identifying dwarf galaxies ($M_{\star}<3\e{9} \ M_{\odot}$) with DEEP2 spectroscopic redshifts $z<1$ in the AEGIS field using the NMBS. We then analyze these sources using \textit{Chandra} X-ray Observatory archival data. Section 3 details the results of our search, while Section 4 discusses the Eddington and AGN fractions that we measure and how these results compare to other studies and simulations. Section 5 is a summary. Throughout, we assume the standard, flat $\Lambda$CDM cosmology with parameters: $H_0 = 68 \ \rm{km \ s}^{-1} \ \rm{Mpc}^{-1}$ and $\Omega_m = 0.3$.

\section{Data \& Source Selection}

\begin{figure}[h]
\epsscale{1}
\centering
\plotone{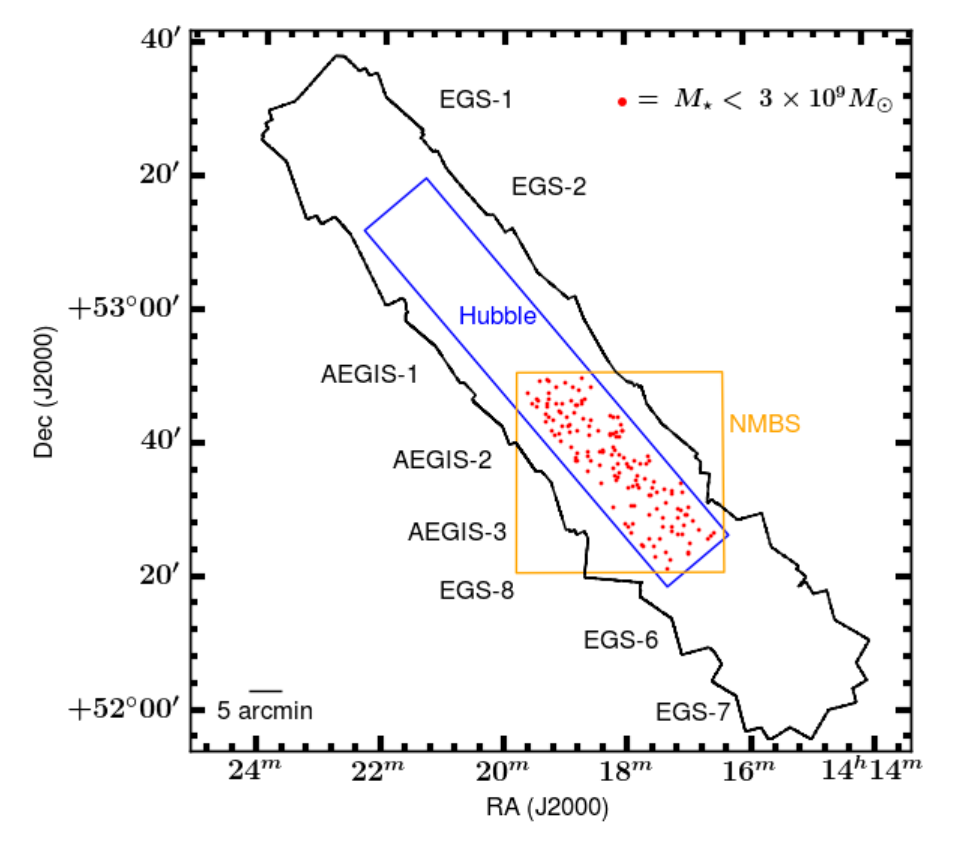}
\caption{Survey region with galaxies selected from the NEWFIRM Medium-Band Survey (NMBS) indicated. The black line outlines the \textit{Chandra} X-ray Observatory AEGIS observations. The blue line corresponds to the \textit{Hubble Space Telescope} observations, and the yellow line shows the region covered by the NMBS. The red circles indicate the positions of the low-mass ($M_{\star}<3\e{9} M_{\odot}$) galaxies with DEEP2 spectroscopic redshifts $z<1$ selected from the NMBS.}
\label{fig-region}
\end{figure}

Our goal is to select dwarf galaxies ($M_{\star} \leq 3\e{9} \ M_{\odot}$) that may harbor central SMBHs. For this, we require deep X-ray data and complementary multi-wavelength data that can provide estimates of the galaxy properties. In addition, it is important to have deep, homogenous spectroscopic redshifts and high spatial resolution imaging. The All-Wavelength Extended Groth Strip International Survey (AEGIS) field has deep X-ray data ($\sim 200-800$ ks) while still providing sufficient volume for statistically significant source numbers and excellent multi-wavelength data. The AEGIS field is centered at $\alpha=14^{\rm h} 17^{\rm m}$, $\delta = +52^{\circ} 30^{\rm m}$ (J2000), and covers an area of $\sim 0.9 \ \rm{deg}^2$ \citep{Davis2002, Davis2007}. The requirement of data from the NMBS, \textit{HST} and \textit{Chandra} within the AEGIS field yields a contiguous overlapping footprint of $\sim 0.1 \ \rm{deg}^2$ (see Figure \ref{fig-region}). We select galaxies within this region as described below. 

\subsection{A Dwarf Galaxy Sample with Robust Stellar Mass Measurements}

\begin{figure}[h]
\centering
\plotone{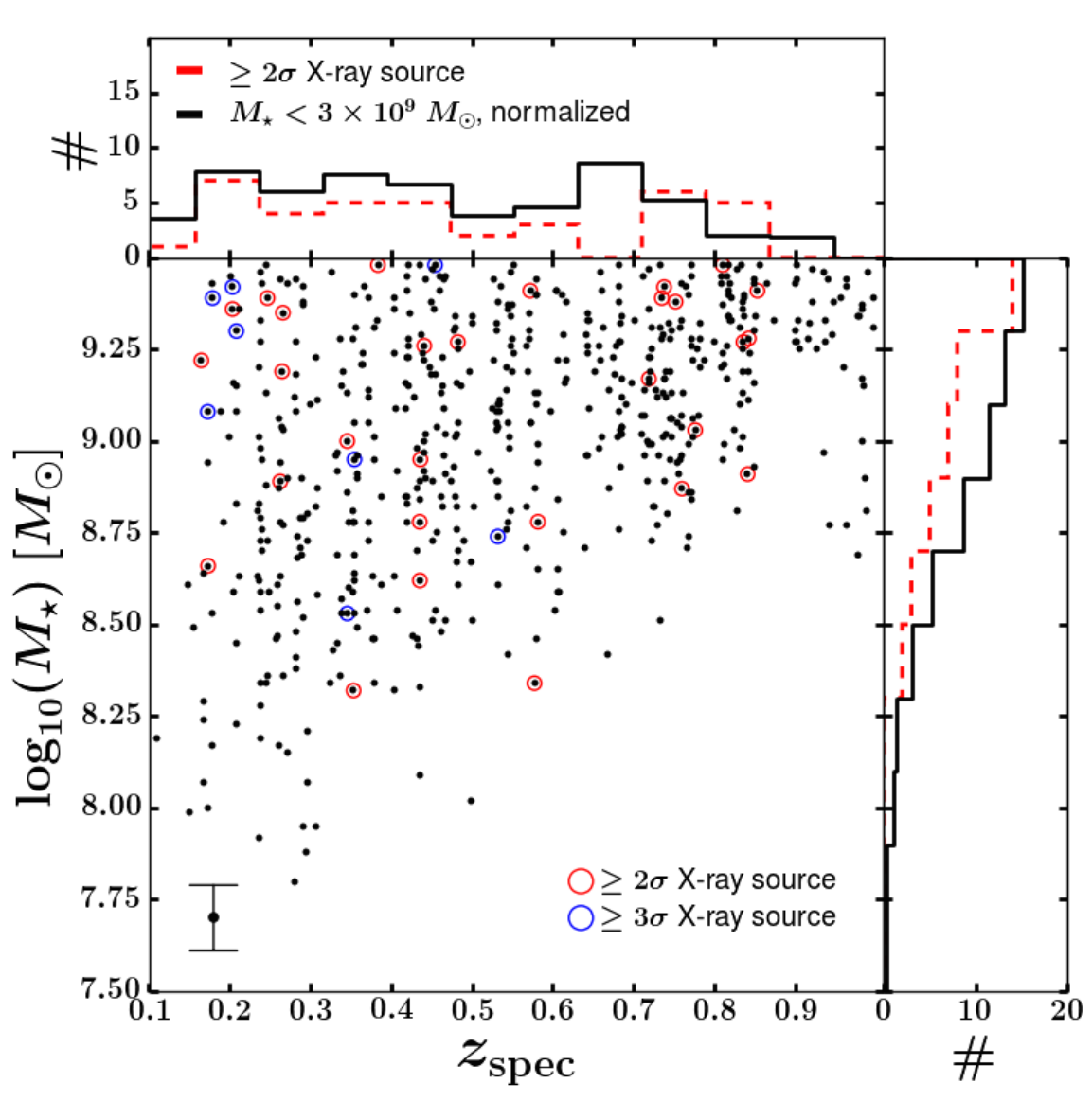}
\caption{Stellar Mass as a function of redshift, with mass and redshift histograms for the dwarf galaxies selected from the NMBS. The black points and lines indicate all dwarf galaxies selected from the NMBS, with the histograms normalized to $10\%$ of their actual values to enable better comparison with the $\geq 2\sigma$ X-ray sources. The red circles and red dotted lines indicate our $\geq 2\sigma$ X-ray sources. The blue circles signify our $\geq 3\sigma$ X-ray sources.}
\label{fig-candidates}
\end{figure}

For reliable stellar masses, we turn to the NEWFIRM Medium-Band Survey \citep[NMBS; see][for a complete description]{Whitaker2011}, which is a near-infrared imaging survey of part of the AEGIS field. Since this survey splits the broad-band $JHK$ bands into $7$ medium-band filters, there is improved modeling of the spectral energy distribution, which leads to more dependable stellar masses \citep{vanDokkum2009}. The $27'.6 \times 27'.6$ NMBS survey region is centered at $\alpha=14^{\rm h} 18^{\rm m} 00^{\rm s}$, $\delta = +52^{\circ} 36^{\rm m} 07^{\rm s}$ (J2000) \citep{vanDokkum2009}. The K-band $90\%$ limiting magnitude is $22.5 \ \rm{AB \ mag}$ and the $50\%$ limiting magnitude is $23.6$ AB mag \citep{Whitaker2011}. The NMBS sources were K-band selected using SExtractor \citep{Bertin1996}, and the stellar population parameters, including star formation rates, were derived with FAST \citep{Kriek2009} using the \cite{Bruzual2003} models and an exponentially declining star formation history. The \cite{Bruzual2003} models use the \cite{Chabrier2003} initial mass function. To ensure more accurate stellar masses, the NMBS used spectroscopic redshifts from the DEEP2 survey, which has a limit of $R < 24.1$ mag, \citep{Davis2002,Steidel2003} when available. For our purposes, accurate stellar masses are paramount, hence we restrict ourselves to the sources with spectroscopic redshifts. In addition, we restrict the galaxies to those with $z_{spec} < 1$ to match the mass sensitivity of the NMBS. Of the $27652$ sources identified by the NMBS, we find $642$ within our mass and redshift ranges that have spectroscopic redshift measurements.

To ensure that all of our sources are dwarf galaxies rather than tidal tails mistakenly identified by the NMBS, we require morphologies of our sources. \textit{HST} surveyed a $0.197 \ \rm{deg}^2$ region of the AEGIS field using the Advanced Camera for Surveys (ACS) \citep{Davis2007}. Using the \textit{HST} data, we discard $37$ of the galaxies within our mass and redshift ranges because they are mergers. We exclude these mergers because of the possibility of inaccurate mass measurements. This leaves us with $605$ sources selected from the NMBS and \textit{HST} data.

The 605 galaxies selected are shown, binned by mass and redshift, in Figure \ref{fig-candidates}. The masses range from $\sim 5\e{7} \ M_{\odot}$ to $3\e{9} \ M_{\odot}$. The star formation rates (SFR) vary from $\sim 3\e{-4}$ to $10\ M_{\odot} \ \rm{yr}^{-1}$, with a median SFR $= 0.24 \ M_{\odot} \ \rm{yr}^{-1}$.

\subsection{X-ray Analysis}

We now identify the dwarf galaxies from our sample that show significant X-ray emission by using the available X-ray data within the AEGIS field. The \textit{Chandra} X-ray Observatory has observed $\sim 0.7 \ \rm{deg}^2$ of AEGIS with the Advanced CCD Imaging Spectrometer (ACIS) as part of XDEEP2 \citep[][see also \citealt{Nandra2005, Laird2009, Nandra2015}]{Goulding2012}. To ensure the most reliable measurements, we apply the latest calibration files using the CIAO software (version 4.7) and CALDB 4.6.5 \citep{Fruscione2006}.

\begin{figure*}[t]
\epsscale{1}
\centering
\plotone{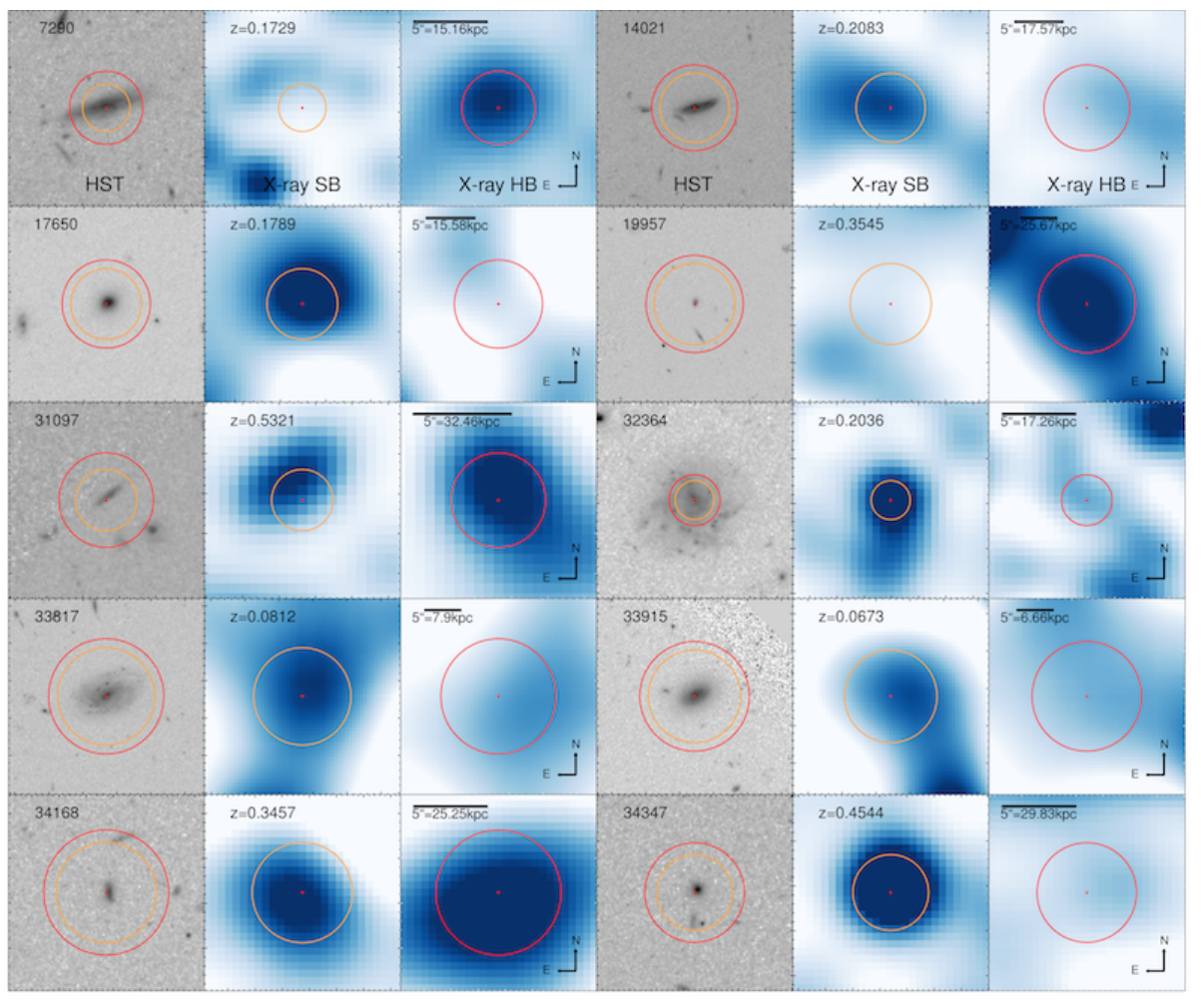}
\caption{\textit{Hubble Space Telescope}(left) images alongside \textit{Chandra} X-ray Observatory soft-band (SB; 0.5 - 2 keV; center) and hard-band (HB; 2 - 7 keV; right) images of our $\geq 3\sigma$ objects. The red x's mark the galaxy position, while the $90\%$ enclosed energy radii are shown in orange (SB) and red (HB).}
\label{fig-composites1}
\end{figure*}

Our X-ray analysis consists of performing forced aperture photometry at the optical positions of our selected galaxies. Some of our selected galaxies have X-ray data from different, overlapping subregions within AEGIS. However, we only use the X-ray data from the region with an aim-point closest to each source. This is because we wish to limit the amount of contamination from diffuse emission, and the \textit{Chandra} point-spread function (PSF) deteriorates with distance from the aim-point. We define the X-ray source region as the $90\%$ enclosed energy radius around the optical source position using the CIAO \texttt{psf} module. This is done for the soft band (SB; $0.5-2$ keV), the hard band (HB; $2-7$ keV), and the full band (FB; $0.5-7$ keV). The $90\%$ enclosed energy radii range from $\sim 1'' - 15''$ for most of our sources. We define the background region to be a square with sides that are at least four times as large as the radius of the source region. We find that $23$ of the $605$ dwarf galaxies selected from the NMBS have a second spatially coincident galaxy within the $90\%$ enclosed energy radius. The X-ray photons associated with these neighbors cannot be reliably disentangled from the target dwarf galaxy, and thus these $23$ sources are discarded from the sample.

Note that our method here differs from that of traditional X-ray catalogs created independently of multi-wavelength source positions. Using X-ray data alone to detect a source necessitates a high cut in X-ray signal-to-noise. However, our forced photometry method allows us to identify X-ray sources that may fall below the typical significance level cut and helps to mitigate the effects of Eddington bias \citep[see][]{Gibson2012}. Since we already know that there is an optical source at a given position, we can accept a lower level of significance and still expect almost no false positives from random background fluctuations. Assuming a false detection probability of $0.27\%$ (a 3-sigma confidence limit), we expect only $\sim 2$ false positives from our $583$ sources (605 sources minus 23 sources with contaminants). This approach is standard for deep field Spitzer data, and WISE catalog construction \citep[e.g.,][]{Lang2014}. In addition, a similar implementation has recently been performed on the CDF-S/N by \cite{Xue2016}.

We determine the flux for each source using the same prescription followed in \cite{Goulding2012}. A count rate to flux conversion is applied assuming a power law spectrum with a photon index of $1.9$, and Galactic $N_{\rm{H}}$ $\sim 1\e{20} \ \rm{cm}^{-2}$ \citep{Kalberla2005}. We calculate the X-ray luminosity for each of our sources using the DEEP2 spectroscopic redshifts, and bring each X-ray luminosity to the rest frame by K-correcting the luminosities using a typical power-law with $\Gamma=1.9$ \citep{Brandt2005}. Finally, we calculate the hardness ratios ($HR$s) for each of our sources, which is given by
\begin{equation}
HR = \frac{H - S}{H+S} \; ,
\end{equation}
where $H$ is the number of HB counts and $S$ is the number of SB counts. We use the Bayesian Estimation of Hardness Ratios code \citep{Park2006} to calculate the $HR$ and its associated uncertainties. $HR$s vary from $-1$ to $1$, with typically more positive values for AGN and more negative values for X-rays produced by stellar processes \citep{Brandt2005}.

For each of our sources, we compute a false detection probability, which gives the probability of a spurious X-ray source being falsely identified as a significant source. In order to compute the false detection probability, we assume that the number of background counts have a Poisson distribution with a mean equal to the number of counts observed in the background region normalized to the area of the source region for each source. We then sample from this distribution $10^6$ times. The false detection probability is the percentage of times that the sampled distribution produced counts greater than the observed source counts. We calculate this probability for the SB, HB, and FB. The false detection probability is used to set the significance of our sources (i.e. a false detection probability $< 4.55\%$ denotes a $\geq 2\sigma$ source). We consider a source to be a $\geq 3\sigma$ or $\geq 2\sigma$ detection if it is significant to that level in any of the three bands. Given the effective area of the ACIS instrument, \emph{Chandra} is most sensitive and has the smallest PSF at soft energies, where the background is also the lowest. In the low-number Poisson regime, there is a further trade-off that more robust detections can be made over the widest possible energy band. Thus, our false detection probabilities are assigned as follows: full-band, soft-band and hard-band. Note that the NMBS and \textit{Chandra} astrometry is matched within $\sim 0\farcs 9$ \citep{Nandra2015}, so we allow our $\geq 3\sigma$ X-ray centroid positions to vary by up to $1''$ before applying the counts to flux conversion.


\section{Results}
There are 151 dwarf galaxies selected from the NMBS with X-ray fluxes above the detection threshold. As described in detail below, we find $10$ sources with $\geq 3\sigma$ detections and $29$ sources with  $\geq 2\sigma$ significant detections (Section 3.1). We give the optical properties of our $\geq 2\sigma$ sources in Table \ref{tab-basicprop}, and their X-ray properties in Table \ref{tab-xprop}. We discuss the nature of these X-ray sources in Section 3.2.

\begin{figure*}[!htb]
\epsscale{1}
\centering
\plotone{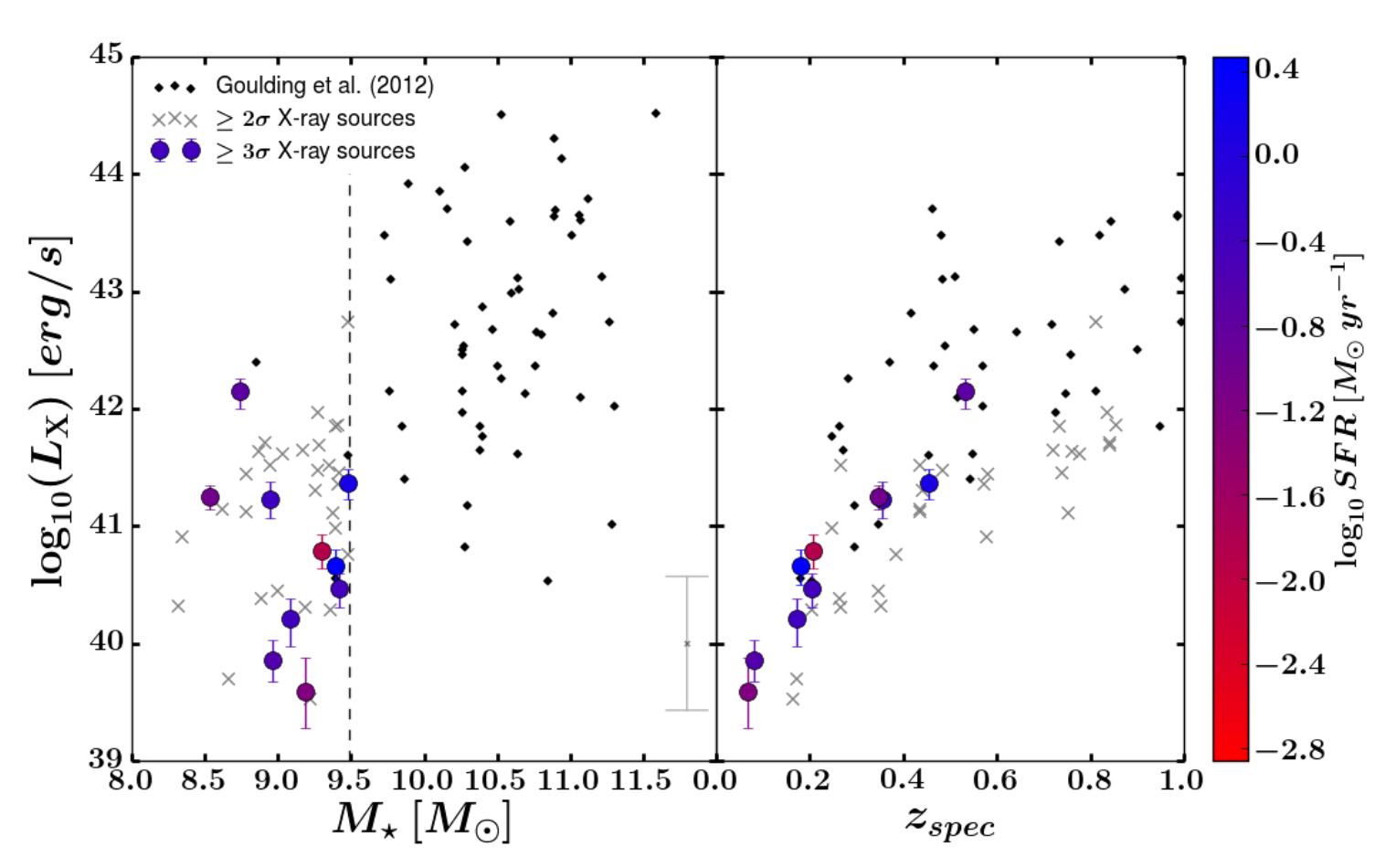}
\caption{$L_{\rm{X}}$ as a function of $M_{\star}$ (\textit{left}) and redshift (\textit{right}) for our $\geq 3\sigma$ X-ray sources (colored points), $\geq 2\sigma$ X-ray sources (grey points), and sources found in \cite{Goulding2012} that correspond to galaxies in the NMBS (black points). The colors correspond to the SFR, as given by the NMBS.  The black, dashed line shows our mass cut ($M_{\star} < 3\e{9} \ M_{\odot}$).}
\label{fig-lummstar}
\end{figure*}

\subsection{X-ray Detections}

We consider any $\geq 3\sigma$ sources (those with false detection probability $<0 .27\%$) to be a robust X-ray detection. These ten sources are shown circled in blue in Figure \ref{fig-candidates} and their properties are given in Table \ref{tab-sigsources}. Their \textit{HST} and \textit{Chandra} X-ray Observatory images are given in Figure \ref{fig-composites1}. Of the ten, seven are significant in the FB and one other band, two are only significant sources in the SB, one is only significant in the HB, and two sources are significant in all three bands. Note that the observed SB and HB correspond to rest-frame $0.75-3$ keV and $3-10$ keV, respectively at $z=0.5$. The redshifts range from $z\approx 0.08-0.53$, although most of the sources are below $z\sim 0.3$. This could be due to the incompleteness of the DEEP2 spectra, which we discuss further in Section 4.2. The rest frame X-ray luminosities of these sources range from $L_{\rm{X}}\sim 4\e{39} \ \rm{erg \ s}^{-1}$ to $2\e{42} \ \rm{erg \ s}^{-1}$. The median hardness ratio for these sources is $-0.09$ and the median SFR is $\sim0.3 \ M_{\odot} \ \rm{yr}^{-1}$. \cite{Goulding2012} performed a blind X-ray search in the AEGIS field, and found three sources that correspond to galaxies in the NMBS within our stellar mass range. Two of those sources, 17650 and 34347, are also identified by our method. Our estimated $L_{\rm{X}}$ agrees with that from \cite{Goulding2012} for both sources. The final source they identified, 31041, was removed from our sample because it is a merging system.

Using the \textit{HST} data, we can glean some basic information about the morphologies of our sources. Most are late-type spirals at a variety of inclinations (see Figure \ref{fig-composites1}). However, one (19957) is irregular, and one at high redshift (34347) is compact with little else about its morphology readily identifiable. A more quantitative analysis of the morphologies (e.g., to determine their bulge-to-disk ratios) is beyond the scope of this paper.


Our $\geq 2\sigma$ sources (those with false detection probability $<4.55\%$) are included as grey x's in Figures \ref{fig-lummstar} and \ref{fig-hr}, and their properties are included in Tables \ref{tab-basicprop} and \ref{tab-xprop}. As described in Section 2.2, because our $\geq 2\sigma$ sources are at known galaxy positions, the vast majority are real sources. We include these sources as they enable us to have a large enough sample to give better statistics on the AGN properties of our sample as a whole (Section 4).

\subsection{Nature of the X-ray Sources}
There are multiple processes that may produce X-ray emission in a star-forming galaxy other than SMBH accretion. High-mass X-ray binaries (HMXBs) are the dominant source of X-rays from star-forming regions \citep[e.g.][]{Fragos2013}. Hot gas associated with supernovae remnants can also emit in the X-ray, although this should be a very small factor for such low star formation rates \citep[see][]{Mineo2012b}. At $L_{\rm{X}}>10^{40} \ \rm{erg \ s}^{-1}$, HMXBs contribute many orders of magnitude more to the luminosity than the low-mass X-ray binaries \citep[see, for example,][]{Lehmer2010,Mineo2012}. Thus, we will mostly consider the contribution from HMXBs, but will also discuss LMXBs and X-ray emitting hot gas from supernovae remnants as AGN contaminants.

\begin{figure}[h]
\epsscale{1}
\centering
\plotone{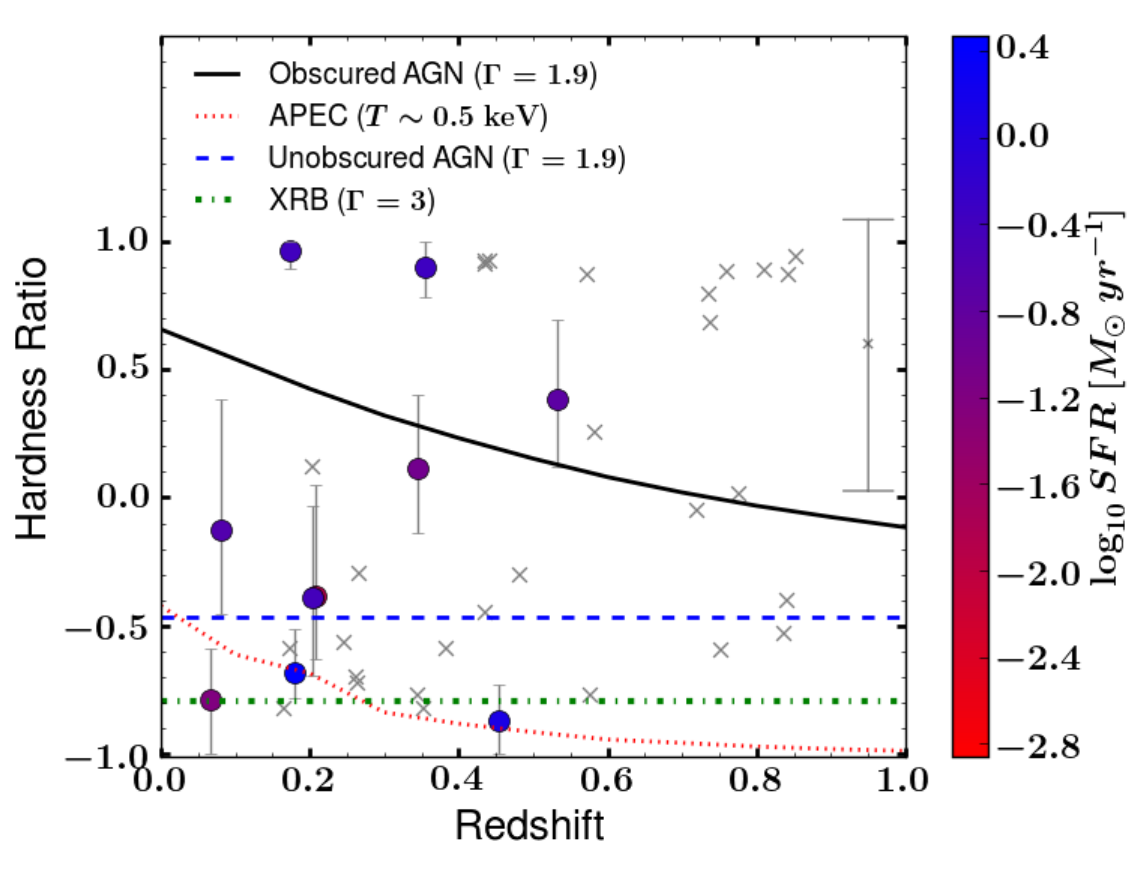}
\caption{Hardness Ratio versus $z$ for $\geq 2\sigma$ sources selected from the NMBS. The Hardness Ratio, $(H-S)/(H+S)$, is calculated using the Bayesian Estimation of Hardness Ratio Code. The colored points are all of the sources selected from the NMBS with $\geq 3\sigma$ X-ray detections, while the grey points indicate the $\geq 2\sigma$ X-ray sources. The colors correspond to the SFR, as given by the NMBS. The various lines give hardness ratios for different models as calculated by the PIMMS online calculator.}
\label{fig-hr}
\end{figure}

Although we cannot measure detailed X-ray spectra with so few counts, we can use hardness ratios as a proxy for the type of X-ray emission we are observing. Obscured AGN (intrinsic $N_{\rm{H}} \gtrsim 1\e{22}\ \rm{cm}^{-2}$)  normally have the highest hardness ratios of $HR > 0$, followed by unobscured AGN with $HR \sim -0.5$. HMXBs and X-ray emitting hot gas associated with supernovae remnants normally have very low hardness ratios of $HR \leq -0.8$ \citep{Brandt2005}. The hardness ratios for our $\geq 2\sigma$ and $\geq 3\sigma$ sources are given in Tables \ref{tab-xprop} \& \ref{tab-sigsources}, and plotted versus the spectroscopic redshift in Figure \ref{fig-hr}. We plot the expected hardness ratio as a function of redshift for an obscured AGN with $\Gamma = 1.9$ and intrinsic $N_{\rm{H}} = 3\e{22}\ \rm{cm}^{-2}$ (black, solid line), an unobscured AGN with $\Gamma = 1.9$ (blue, dashed line), an HMXB with $\Gamma = 3$ (green, dotted-dashed line), and hot gas, for which we use the APEC model provided by Sherpa, at $\sim 0.5$ keV (red, dotted line). We calculate each of these with the PIMMS online calculator\footnote{http://cxc.harvard.edu/toolkit/pimms.jsp} and allow for a constant galactic $N_{\rm{H}} = 1\e{20} \ \rm{cm}^{-2}$. As shown in the figure, three of our sources (7290, 19957, \& 31097) are clearly obscured AGN as they fall well above the track shown for the model unobscured AGN. Most of our other sources are consistent with being either an obscured or unobscured AGN. Our three softest sources (17650, 33915, \& 34347) appear inconsistent with either AGN model. These low $HR$s could be consistent with AGN if these AGN were extremely obscured (e.g. reflection dominated X-ray spectra produced in the presence of Compton thick absorption). On the other hand, these $HR$s are consistent with the normal X-ray emission from star formation. This second scenario seems more likely for two of these sources (17650 \& 34347) considering their SFRs of $0.47  \ M_{\odot} \ \rm{yr}^{-1}$ and $0.11 \ M_{\odot} \ \rm{yr}^{-1}$, which are among the highest of our $\geq 3\sigma$ sources. Thus, based on $HR$s, we conclude that at least seven of our ten sources are likely AGN powered.

\begin{figure}[h]
\epsscale{1}
\centering
\plotone{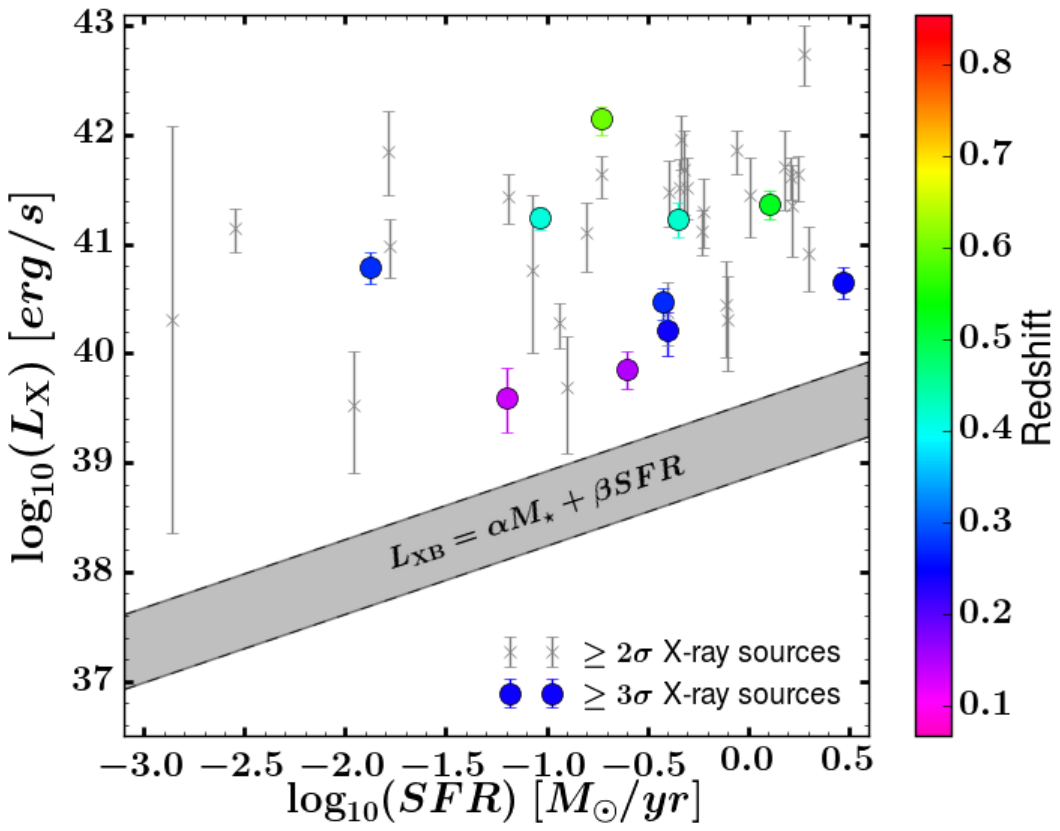}
\caption{X-ray luminosity as a function of star formation rate for the NMBS galaxies with $\geq 2\sigma$ X-ray detections. The shaded grey region indicates the $68\%$ confidence interval for $L_{\rm{XRB}}$. The colored points represent the measured X-ray luminosity in the full-band for our $\geq 3\sigma$ sources. The colors correspond to the redshift of each source. The grey points correspond to our $\geq 2\sigma$ sources. Note that for most of these sources, $L_{\rm{X,min}}$ is still well above $L_{\rm{XRB}}$.}
\label{fig-sfr}
\end{figure}

Now, we consider the possible contamination to the measured $L_{\rm{X}}$ due to HMXBs and LMXBs specifically. To quantify this contamination, we assign each source an X-ray binary (XRB) probability (i.e. the probability that the X-ray luminosity observed can be attributed to HMXBs or LMXBs). We assume Gaussian statistics, setting our mean to the observed X-ray luminosity in the full-band (FB), $L_{\rm{X}}$\footnote{The Gaussian we use is: $\frac{1}{\sigma \sqrt{2\pi}} e^{-{1\over2}\left(\frac{x-\mu}{\sigma}\right)^2}$, where $\mu$ is the observed X-ray luminosity in the FB and $\sigma$ is the error in our observed X-ray luminosity, which we calculate from the observed counts using Poisson statistics.} and comparing it to the estimated X-ray luminosity due to XRBs, $L_{\rm{XRB}}$. We obtain $L_{\rm{XRB}}$ using the star formation rate (SFR) given by the NMBS and the relation given by \cite{Lehmer2010}:
\begin{equation} \label{eq-hmxb}
L_{\rm{XRB}} = \alpha M_{\star} + \beta SFR \; ,
\end{equation}
where $\alpha = (9.05\pm0.37)\e{28} \ \rm{erg \ s}^{-1} \ M_{\odot}^{-1}$ and $\beta = (1.62\pm0.22)\e{39} \ \rm{erg \ s}^{-1} \ (M_{\odot}/\rm{yr})^{-1}$. This relation gives the luminosity in the $2-10$ keV energy range at redshift $z=0$. However, at the typical redshift of our sources, rest-frame 2-10 keV corresponds to observed 0.3-7 keV, which is approximately our FB. The predicted $L_{\rm{XRB}}$ for each of our sources is given in Table \ref{tab-basicprop}. The $68\%$ confidence interval for $L_{\rm{XRB}}$ of each of our sources is shown in Figure \ref{fig-sfr}. The colored points are the measured $L_{\rm{X}}$ in the FB for each of our $\geq 3\sigma$ sources, where the colors correspond to the redshift. The grey points give the measured $L_{\rm{X}}$ in the FB for all of our $\geq 2\sigma$ sources. Eight of our $\geq 3\sigma$ sources are more than $3 \sigma$ away from the expected $L_{\rm{X}}$ from star formation.

Since high-luminosity X-ray sources will only be populated stochastically at these SFRs $< 1 M_{\odot} \ \rm{yr}^{-1}$ \citep{Gilfanov2004}, we also calculate the total number of luminous X-ray binaries expected in the entire sample, including any ultra-luminous X-ray sources (ULXs), which are X-ray sources with $L_{\rm{X}}>10^{39}\ \rm{erg \ s}^{-1}$ associated with star-forming regions. We take the summed SFR for our galaxies with $L_{\rm{X}} \geq 10^{40} \ \rm{erg \ s}^{-1}$, and integrate the best-fit power law for the luminosity distribution of the HMXBs given by \cite{Mineo2012} over the range of luminosities we measured. We find that we should expect $\lesssim 3$ HMXBs/ULXs with $L_{\rm{X}} \geq 10^{40} \ \rm{erg \ s}^{-1}$ in our entire sample. In line with our hardness ratio predictions, the distribution of $L_{\rm{X}}$ (given the SFRs in our sample) strongly suggests that at least seven of our ten sources are powered by AGN in the X-ray. We should stress that the X-ray binary luminosity function is still largely unknown - it is currently unclear how to include second order factors, like metallicity. There is some evidence that the X-ray binary luminosity function depends on metallicity, and is higher for metal-poor systems \citep{Basu2016}. However, this metallicity dependence is still not well known, so we do not attempt to correct for this effect here.

\begin{figure}[t]
\epsscale{1}
\centering
\plotone{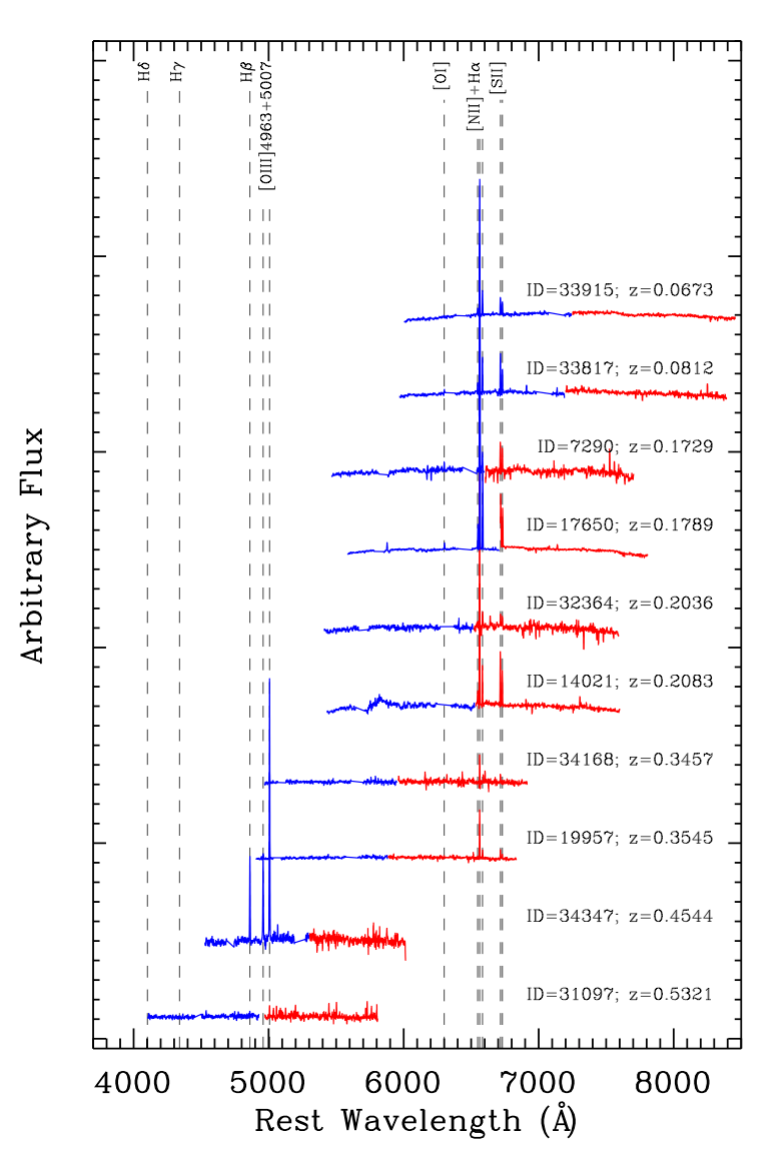}
\caption{DEEP2 spectra with important emission lines marked for our $\geq 3\sigma$ X-ray sources. Note that object 31097 has no emission lines because the lines all fall in detector gaps.}
\label{fig-deep2spec}
\end{figure}

\begin{figure}[h]
\epsscale{1}
\centering
\plotone{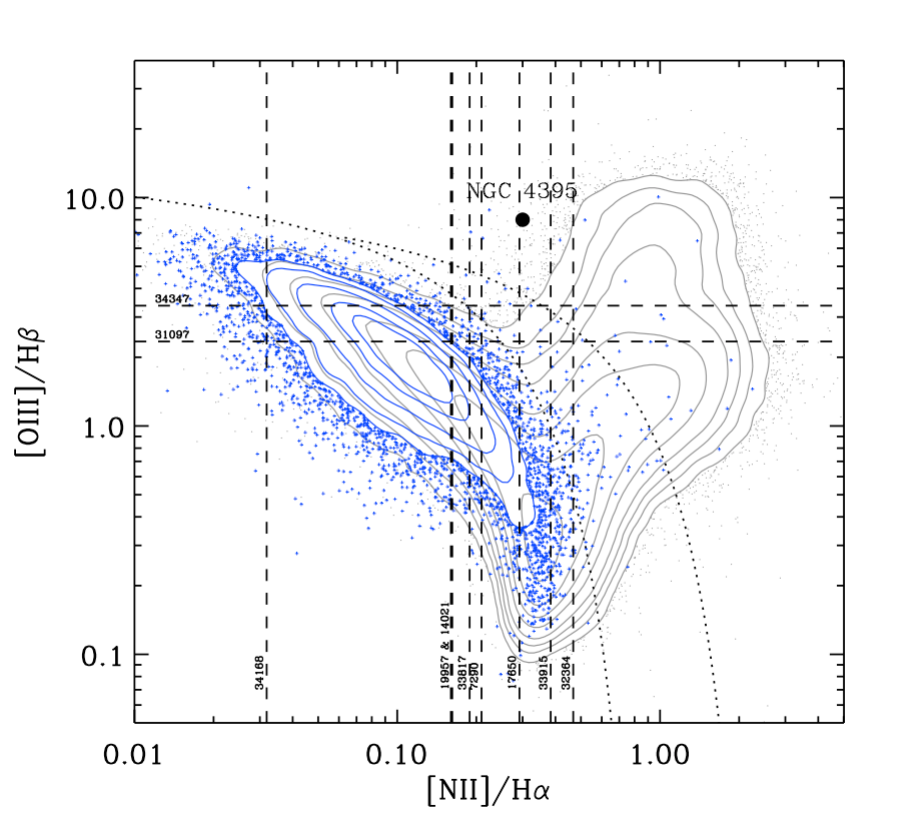}
\caption{A Baldwin, Phillips \& Terlevich (BPT) diagram \citep{Baldwin1981}. The dashed lines indicate the line ratios for our $\geq 3\sigma$ sources. The blue contours are given by all of the dwarf galaxies in SDSS, while the grey contours are all of the AGN in SDSS. The position of NGC 4395 is given as a reference. Note that our sources, as a whole, will most likely lie between the positions of the dwarf galaxies in SDSS and the AGNs in SDSS. The dotted lines give the boundary between starburst galaxies and AGN as found by \cite{Kauffmann2003a} (lower line) and \cite{Kewley2001} (upper line).}
\label{fig-bpt}
\end{figure}

Another way of diagnosing the source of X-ray emission is to use the DEEP2 spectra to plot our sources on a Baldwin, Phillips \& Terlevich (BPT) diagram \citep{Baldwin1981}. These diagrams use line ratios of strong optical emission lines to categorize the source of photoionization in galaxies. Normal star-forming galaxies occupy the left-hand locus of the BPT diagram (see Figure \ref{fig-bpt}), with lower-metallicity sources typically having lower [N II]/H$\alpha$ and higher [O III]/H$\beta$ \citep[e.g.,][]{Moustakas2006}. AGN appear in the upper-right of the diagram, while AGN in dwarf galaxies, which also have low metallicity, tend to fall to the left of the typical AGN sequence due to their lower-than-average [N II]/H$\alpha$ \citep[e.g., NGC4395;][]{Groves2006,Ludwig2009}.

The DEEP2 spectra for our $\geq 3\sigma$ sources are given in Figure \ref{fig-deep2spec}. With the exception of $31097$, which was at the exact redshift where all of the necessary emission lines were in detector gaps, all of our sources exhibit strong [N II], [O III], H$\alpha$, or H$\beta$ emission lines. Due to the coverage of DEEP2, none of the spectra include all of the lines needed to properly place the sources in a BPT diagram. However, we can consider the sample as an ensemble by plotting the ratios we do have for each source (either [N II]/H$\alpha$ or [O III]/H$\beta$). The intersection of these lines places the sample in a spot on the BPT diagram ([N II]/H$\alpha$ $\sim 0.15-0.35$, [O III]/H$\beta$ $\sim 2-3$) between normal SDSS dwarf galaxies and AGN within SDSS (see Figure \ref{fig-bpt}). This is consistent with the low-metallicity AGN described by \cite{Groves2006}. Thus, the BPT diagram indirectly argues for AGN powering in our sources as well.

Therefore, the combination of hardness ratios, $L_{\rm{X}}$/SFR, and optical line ratios together strongly imply that our sample is dominated by AGN.

\section{AGN Properties} 
As mentioned in Section 1, the SMBH occupation fraction in dwarf galaxies can be an important constraint on possible SMBH formation mechanisms. When we use accretion to find supermassive black holes, there is a natural degeneracy between the distribution of black hole mass and the distribution of accretion rates onto the black holes \citep[e.g.,][]{Miller2015}. It is well beyond the scope of this work to attempt to disentangle these two, but we do present active black hole fractions and compare with prior observational and theoretical work.

\subsection{Black Hole Masses \& Eddington Ratios}
Eddington ratios are a useful tool for comparing AGN fractions from different studies; however, the masses of our black holes, $M_{\rm{BH}}$, are very uncertain. As mentioned in Section 1, we cannot measure the black hole masses $M_{\rm{BH}}$ using dynamical methods. We also cannot use the $M_{\rm{BH}}-\sigma_{\star}$ relation to infer $M_{\rm{BH}}$ because we do not have $\sigma_{\star}$ measurements. In addition, the $M_{\rm{BH}}-\sigma_{\star}$ relation has not been well-measured in this low-mass regime \citep[e.g.][]{Barth2004, Dong2012}, and may not hold for late-type spiral galaxies \citep[e.g.][]{Greene2010}. 

For lack of more detailed data, we simply assume that $M_{\rm{BH}}$ scales with $M_{\star}$. This scaling is seen \cite[e.g.,][]{Reines2015}, albeit with considerable scatter. To bracket this range, we take two extremes of the galaxy population at low mass: NGC 4395 and M32. NGC 4395 is a low-mass bulgeless spiral galaxy with a measured central SMBH mass of $M_{\rm{BH}} = 4^{+8}_{-3}\e{5} M_{\odot}$ \citep{denBrok2015}. M32 is a dwarf elliptical galaxy of a similar mass as NGC 4395, but with a measured central SMBH mass of $M_{\rm{BH}} = (2.4 \pm 1.0) \e{6} M_{\odot}$ \citep{van-den-Bosch2010}. We use NGC 4395 and M32 to bracket the range of $M_{\rm{BH}}/M_{\star}$. For NGC 4395,  $M_{\rm{BH}}/M_{\star, 4395} = 3.5\e{-4}$ \citep{Reines2015}, while for M32 this is $M_{\rm{BH}}/M_{\star, M32}=8\e{-4}$ \citep{van-den-Bosch2010}. We derive a corresponding average black hole mass range of $M_{\rm{BH}} \sim 5.5\e{5}-1.3\e{6} \ M_{\odot}$ for both our $\geq 2\sigma$ and $\geq 3\sigma$ sources. To find the Eddington ratios, we assume that $L_{\rm{X}} = 0.1\times L_{\rm{bol}}$ \citep{Marconi2004}. The range of average Eddington ratios, $L_{\rm{bol}}/L_{\rm{Edd}}$, for our $\geq 3\sigma$ sources is $3-7\%$, and $3-6\%$ for our $\geq 2\sigma$ sources.

\subsection{The Observed Fraction of AGN in Dwarf Galaxies}
This paper represents the first attempt to quantify the AGN fraction beyond $z\sim0.4$ in $\sim10^9 \ M_{\odot}$ galaxies. At the very least, the AGN fraction represents a lower limit on the fraction of dwarf galaxies that host massive black holes. As we will show, we are at the limits of the capabilities of both the optical and X-ray surveys, which makes quantifying our incompleteness challenging.

Let us start with our X-ray completeness. Examination of Figure \ref{fig-lummstar} shows that for redshifts $0.2 < z < 0.8$, we are incomplete below an X-ray luminosity of $L_{\rm{X}} \sim 10^{41}\  \rm{erg \ s}^{-1}$. Thus, for calculating our AGN fraction we will only consider sources with $L_{\rm{X}}$ greater than this limit across our redshift range. Our optical limit, on the other hand, has a more complicated selection function. Our stellar masses are derived from the NMBS, whose point-source detection limit is $m_{\rm{K}} = 23.2 \ \rm{AB}$. However, we also require a spectroscopic redshift from DEEP2, which introduces a known incompleteness for faint red galaxies with $z>0.7$ \citep{Newman2013}. We examine the ratio of the NMBS to DEEP2 sources in redshift bins, and find that this ratio remains constant out to $z\sim0.8$ if we restrict our attention to galaxies with $M_{\star} > 10^9 \  M_{\odot}$. However, to better compare with \cite{Aird2012} (see Section 4.3), we we will only consider up to $z<0.6$. Thus, for calculating an AGN fraction, we also restrict our attention to a mass and redshift range of $10^9 \ M_{\odot} \leq M_{\star} \leq 3\e{9} \ M_{\odot}$ and $0.1<z<0.6$.

\begin{figure}[h]
\epsscale{1}
\centering
\plotone{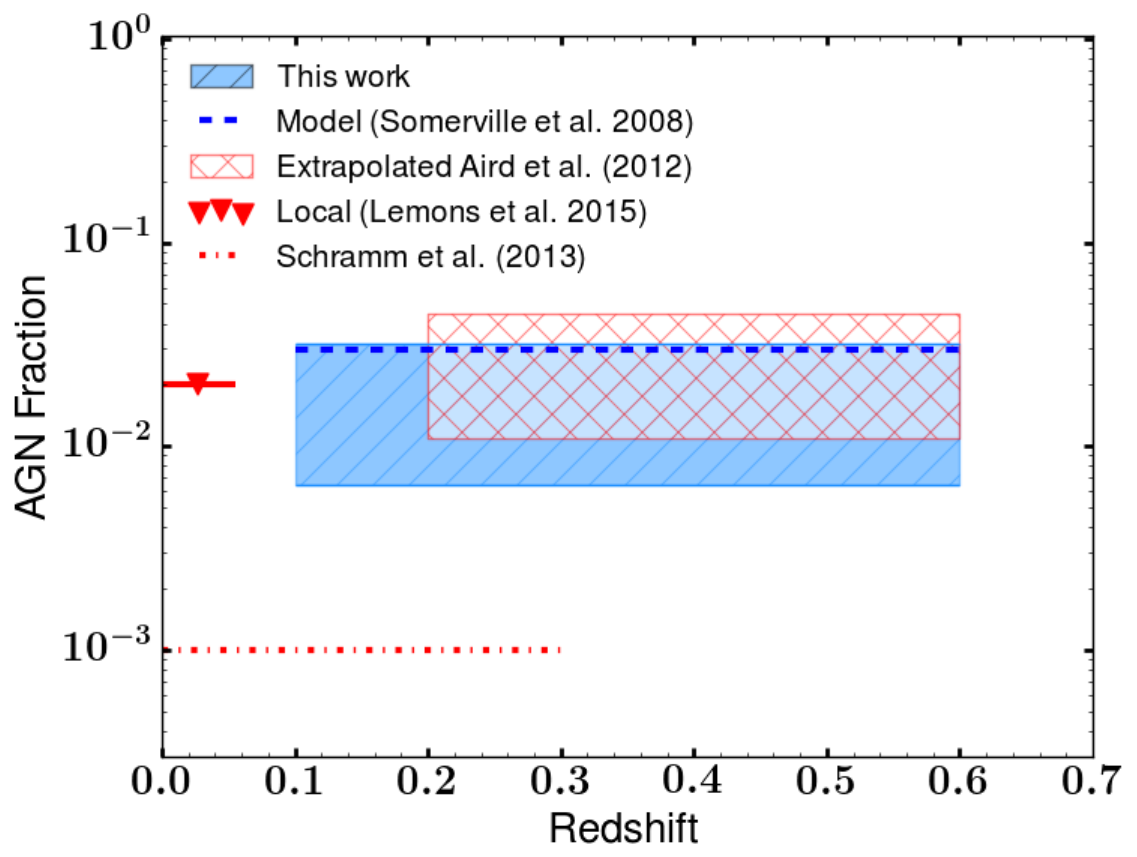}
\caption{AGN Fraction as a function of redshift. The blue, striped region gives our measured AGN fraction for our $\geq 2\sigma$ and $\geq 3\sigma$ sources with $10^9 \ M_{\odot} \leq M_{\star} \leq 3\e{9} \ M_{\odot}$, $0.1<z<0.6$, and $L_{\rm{X}} \geq 10^{41} \ \rm{erg \ s}^{-1}$. The blue dashed line gives the AGN fraction found using semi-analytic models and applying the same luminosity and mass cuts. The red, hashed region the AGN fraction range found by extrapolating from \cite{Aird2012} for $10^9 \ M_{\odot} \leq M_{\star} \leq 3\e{9} \ M_{\odot}$ and $0.2<z<0.6$. The red triangle gives the \cite{Lemons2015} AGN fraction.}
\label{fig-activefrac}
\end{figure}

The AGN fraction, $f_{AGN}$, for our sources within these limits is given by
\begin{equation}
f_{AGN} = \sum^N_{i=0} \frac{(1-P^i_{\rm{false}})(1-P^i_{\rm{XRB}})}{N_{\rm{total}}} \; ,
\end{equation}
where $P_{\rm{false}}$ is the false detection probability, $P_{\rm{XRB}}$ is the XRB probability (see Section 3.2), and $N_{\rm{total}} = 154$ is the total number of galaxies analyzed with mass $10^9 \ M_{\odot} \leq M_{\star} \leq 3\e{9} \ M_{\odot}$ and $0.1<z<0.6$. All of our $\geq3\sigma$ sources with luminosity above the incomplete luminosity limit, $L_{\rm{X}} > 10^{41} \ \rm{erg \ s}^{-1}$, are included in this calculation. This leaves us with one $\geq3\sigma$ source (34347), which would indicate an AGN fraction for our $\geq 3\sigma$ sources of $0.6\%$. However, because of the large uncertainties in this calculation, we can use the $\geq 2\sigma$ sources to provide an upper bound to the AGN fraction. In this case, we have $5$ sources within the mass, redshift, and luminosity limits. This gives an AGN fraction for our $\geq 2\sigma$ sources of $3\%$. Thus, we find an AGN fraction of $\sim 0.6-3\%$. This range of fractions are shown in Figure \ref{fig-activefrac}. 

\subsection{Comparison of Observed \& Theoretical AGN Fractions in Dwarf Galaxies}
Several studies have measured AGN fractions for dwarf galaxies in the local universe. It is highly non-trivial to compare our fractions with these other studies, given the different techniques, depths, and redshifts of the different surveys. In an attempt to provide somewhat meaningful comparisons, we here focus on X-ray surveys. Note that most of these studies cross-correlate optical positions of galaxies with an X-ray catalogue rather than performing forced X-ray photometry at the optical positions as we do in this paper. Figure \ref{fig-activefrac} and Table 4 give summaries of these AGN fractions.

\cite{Miller2015} study $\sim 200$ early-type dwarf galaxies in the local Universe selected optically from \textit{HST} imaging for the AGN Multi-wavelength Survey of Early-type galaxies (AMUSE) surveys \citep[see][for details on these surveys]{Gallo2008,Miller2012}. They then use the \textit{Chandra} X-ray Observatory data from the AMUSE survey and Bayesian statistical techniques to study the AGN fraction in galaxies with $M_{\star}<10^{10} \ M_{\odot}$. \cite{Miller2015} report a SMBH occupation fraction of $>20\%$ down to $L/L_{\rm{Edd}} \sim 10^{-4}$ for $M_{\star}<10^{10} \ M_{\odot}$. Due to our average sample redshift, we do not probe this low Eddington regime here. \cite{Lemons2015} also focus on the very nearby universe in their survey of dwarf galaxies. They select a sample of $\sim 44,000$ dwarf galaxies from the NASA-Sloan atlas\footnote{http://www.nsatlas.org/} with $z<0.055$. \cite{Lemons2015} then cross match these sources to the X-ray sources within the \emph{Chandra} Source Catalog. Although they do not give the number of galaxies in the NASA-Sloan atlas that are within the \emph{Chandra} Source Catalog area, we can estimate this number using the relative areas of each of these fields. The SDSS DR7 footprint is $11,663 \ \rm{deg}^2$ \citep{Abazajian2009}, and the \emph{Chandra} Source Catalog SDSS observations have an area of $130 \ \rm{deg}^2$ \citep{Goulding2014}. So, we estimate that of the $\sim 44,000$ dwarf galaxies in the NASA-Sloan atlas \cite{Lemons2015} selected, $\sim 491$ dwarf galaxies are within the \emph{Chandra} Source Catalog area. \cite{Lemons2015} find $19$ dwarf galaxies with hard X-ray detections, although they argue that at most $10$ of these are indicative of a central SMBH. This gives an estimated AGN fraction of $\sim 2\%$ for $M_{\star} < 3\times10^{9} \ M_{\odot}$ and $z<0.055$.  At somewhat higher redshifts, \cite{Schramm2013} optically select $\sim 5200$ dwarf galaxies with $M_{\star} < 3\times10^{9} \ M_{\odot}$ at $z<1$ using \textit{HST}/ACS imaging from the Galaxy Evolution from Morphologies and SEDs (GEMS) survey \citep{Rix2004}.  They then use multi-wavelength catalogs to match these galaxies to X-ray sources in the \emph{Chandra} Deep Field-South. \cite{Schramm2013} find $27$ X-ray detected galaxies among their $\sim 5200$ dwarf galaxy sample up to $z<1$. However, they then focus on the sample with $z<0.3$, optical spectra, and a high ratio of X-ray to optical luminosity, which leaves them with a parent sample of $\sim 2100$ galaxies and $3$ X-ray detected galaxies. They argue that this sample of $3$ galaxies all contain AGN by examining their H$\alpha$ or UV fluxes. Ignoring selection effects, we calculate an AGN fraction of $\sim 0.1\%$. However, \cite{Schramm2013} does not discuss the incompleteness of the sample.

Perhaps the most straightforward comparison is made with the X-ray study of \cite{Aird2012}. \cite{Aird2012} select $25,000$ galaxies that are within fields with X-ray data from the PRIMUS survey \citep{Aird2012}. They identify 242 AGN with $L_{\rm{X}} > 10^{42} \ \rm{erg \ s}^{-1}$ out of this sample using archival \emph{Chandra} and \emph{XMM-Newton} X-ray data.  They report an AGN fraction of $\sim 1\%$ up to $z\sim 1$ but for more massive galaxies than we consider here. While \cite{Aird2012} focus on different mass ranges than we do here, we can compare to their study because of the central result of that paper: \cite{Aird2012} show that AGN accretion rate distributions do not depend upon the host galaxy stellar mass. They report a universal Eddington ratio for $M_{\star}>3\e{9} \ M_{\odot}$ and $L_{\rm{X}} > 10^{42} \ \rm{erg \ s}^{-1}$ in the redshift range\footnote{Note that while \cite{Aird2012} did search to $z<1$, there were no objects in their lowest mass bin at the higher redshifts.} $z=0.2-0.6$. Given that the AGN accretion rate distribution is independent of the host galaxy stellar mass, we can extrapolate these probability densities to the mass and luminosity ranges we study here. For this linear extrapolation, we use Aird's fit to the figure of active fraction (i.e. probability density of AGN with a given luminosity, $L_{\rm{X}}$ in a galaxy of stellar mass, $M_{\star}$) as a function of AGN X-ray luminosity and stellar mass \citep[Top panels of Figure 4 in][]{Aird2012}. By virtue of the universal accretion rate distribution, we adopt their measured slope (i.e. $P(L_{\rm{X}} | M_{\star})$ per $\log L_{\rm{X}} \ / \ [\rm{erg \ s}^{-1}]$) of $\alpha = -0.8$ \citep{Aird2012}. We then use a least squares fit to find the value for the AGN fraction at the `intercept' ($L_{\rm{X}} = 10^{43} \ \rm{erg \ s}^{-1}$) for $M_{\star}= 10^9 \ M_{\odot}$, which gives us an intercept of $b =  -3.3 \pm 0.3$. Using these values, we calculate the extrapolated AGN fraction for $z=0.2-0.6$ and $M_{\star} = 10^9 \ M_{\odot}$ (i.e. $P(L_{\rm{X}} | M_{\star}= 10^9 \ M_{\odot}, z=0.2-0.6)$) to be $1.1-4.5\%$. This is remarkably consistent with our own results of $0.6-3\%$ for $10^9 \ M_{\odot} \leq M_{\star} \leq 3\e{9} \ M_{\odot}$ and $0.1<z<0.6$ (see Figure \ref{fig-activefrac}). 





We also compare the fractions of detected active black holes with the predictions of a semi-analytic model (SAM) of the joint formation and evolution of galaxies and SMBHs from \cite{Somerville2008} \citep[see also][]{Hirschmann2012}. SAMs are useful tools because they allow us to model a statistical sample down to very low masses, such as those relevant in this paper. The \cite{Somerville2008} models reproduce dwarf galaxy properties well, and manage to match the AGN luminosity function at higher luminosities \citep{Somerville2012}. Thus, it is interesting to ask whether their AGN fractions are similar to ours, with the eventual hope to test their assumptions about seeding and accretion.

The SAM is based on cosmological merger trees and contains recipes for the usual processes included in models of galaxy formation, including gas accretion and cooling, star formation, and stellar feedback. In this model, each top-level halo is seeded with a $10^4 \ M_{\odot}$ black hole. As shown in \cite{Hirschmann2012}, these seed masses are required to make the luminous AGN observed at $z>5$ without resorting to super-Eddington accretion. Black hole accretion is triggered by mergers with mass ratio greater than 1:10, which is based on the results of hydrodynamic simulations of galaxy mergers \citep{Hopkins2006, Hopkins2007}. The black hole growth is self-regulated by radiative feedback from the accreting black hole, resulting in a power-law decline of accretion rates as described in \cite{Hopkins2006}. Although we consider a SAM in which black hole accretion is triggered only by mergers, in reality, secular processes and internal disk instabilities likely contribute \citep[see][]{Hirschmann2012}. While we exclude visible mergers from our survey, we show in Section 5 that including these mergers would not result in an appreciable change to our results. Moreover, many AGN hosts in the models would not be visually identified as a merger. Many hosts at the accretion levels relevant here had a merger long enough ago that the signatures would no longer be easy to recognize, or were triggered by much smaller objects that would also be difficult to identify in surveys at the present depth.


We select galaxies from a mock AEGIS lightcone using the same stellar mass criteria and redshift range as for the observed sample. Our model predicts the accretion rate onto the black hole, and we convert this to an X-ray luminosity using a conversion of rest mass to energy of 0.1, and the bolometric correction to the hard X-ray band from \cite{Marconi2004}. We do not account for obscuration of the AGN radiation. We then compute the AGN fraction by counting the fraction of galaxies in the relevant stellar mass and redshift bins with $L_{\rm{X}} \geq 10^{41} \ \rm{erg \ s^{-1}}$. 

In this way, we find a detected AGN fraction for galaxies with stellar mass $10^9 \ M_{\odot} \leq M_{\star} \leq 3\e{9} \ M_{\odot}$ and in the redshift range $0.1\leq z \leq 0.6$ of $3.0\%$, which is encouragingly close to the observational results presented here (see Figure \ref{fig-activefrac}). The median Eddington ratio for the active black holes with $L_{\rm{X}}$ greater than the detection limit is $3\%$, while the mean for the same sample is $9\%$. We have also performed the exercise of adding the X-ray flux expected from high-mass X-ray binaries using Equation \ref{eq-hmxb} at the given stellar masses and SFRs. We find that only a few additional galaxies would be above the adopted X-ray detection limit due to the added contribution to the X-ray flux from HMXBs, resulting in a negligible change in our predicted AGN fractions. This is consistent with the results shown in Figure \ref{fig-sfr}. 

While the SAM results agree surprisingly well with the observational results, it is important to understand that significant degeneracy still remains between the Eddington ratio distribution and the occupation fraction. In principle, the scenario that our SAM tests - that of high seed mass and overall low accretion rates - may be correct, in accord with direct gas cloud collapse models \citep[e.g.][]{Eisenstein1995, Regan2009, Mayer2015, Volonteri2010} or formation in nuclear clusters \citep[e.g.][]{Loeb1994, Goswami2012, Katz2015, Stone2016}. On the other hand, there may be a low seed mass and high accretion rates, which agree with Population III star collapse models \citep[e.g.][]{Madau2001, Johnson2012, Becerra2015}. And of course, the truth could be some hybrid of these multiple scenarios. However, the surprising agreement between our observations, the SAM, and the extrapolation from \cite{Aird2012} may suggest that the universal accretion rate distribution does extend to dwarf galaxies. If so, it is most likely that the occupation fraction is of order unity \citep{Miller2015, Plotkin2016}. We are hopeful that more observational constraints, perhaps from JWST \citep[e.g.][]{Windhorst2009}, will help inform these different scenarios. In addition, further studies similar to our own that focus on a wider mass range of galaxies would be beneficial in truly discerning the AGN fraction as a function of mass. This, combined with the observations from the next-generation telescopes, should allow us to constrain the seeding mechanisms further.

We also use the SAM to predict an AGN fraction for the same mass and X-ray luminosity limits in the redshift range $z = 1-1.5$. We find an AGN fraction in this range of $9.4\%$ and a mean Eddington ratio of $10.3\%$ for the active black holes with X-ray luminosities above the detection limit. This predicted AGN fraction is a bit higher than the one we measure at lower redshifts, which could indicate a rise in AGN activity in dwarf galaxies similar to that in more massive galaxies at higher redshifts. This may be measurable in the near future by next generation surveys.

\section{Summary and Future Work}
Here, we study a sample of $605$ dwarf galaxies with the aim of finding AGN and constraining the AGN fraction in these galaxies. First, we identify dwarf galaxies ($M_{\star}<3\e{9} M_{\odot}$) with redshift $z<1$ using the NEWFIRM Medium-Band Survey, along with \textit{HST} observations and DEEP2 spectra. We then perform aperture photometry using available {\it Chandra} ACIS-I data at the optical positions of the dwarf galaxies. We identify 10 AGN with $\geq 3\sigma$ certainty and $29$ AGN with $\geq 2\sigma$ certainty out of our sample of $605$ galaxies. It is difficult to directly separate the emission of high-mass X-ray binaries (HMXBs) from low-luminosity AGN in these galaxies, but we use several different techniques to argue that the majority of our sources are AGN. Most of the hardness ratios of our sources are consistent with AGN rather than HMXBs or X-ray emitting hot gas associated with supernovae remnants, which both have more negative hardness ratios than those we observed. In addition, the $L_{\rm{X}}$ we measure is much higher than the estimated X-ray luminosity due to star formation. Also, the number of HMXBs/ULXs in our sample with X-ray luminosities as high as those observed is estimated to be very low. Furthermore, by plotting the optical line ratios found in the DEEP2 spectra of our sources on a BPT diagram, we find that they tend to occupy a region consistent with other AGN. 

Finally, we use our results to calculate an AGN fraction for dwarf galaxies - the first time this has been directly calculated beyond the local universe. We find a fraction of $0.6-3\%$ for galaxies in the stellar mass range $10^9 \ M_{\odot} \leq M_{\star} \leq 3\e{9} \ M_{\odot}$ and redshift range $0.1\leq z \leq 0.6$. This agrees well with other studies, like \cite{Lemons2015} and \cite{Schramm2013}, that find the AGN fraction for dwarf galaxies in the nearby universe. It is also consistent with the \cite{Aird2012} universal accretion rate distribution. We also compared our results to semi-analytic models. These models seed all top-level halos with $10^4 \ M_{\odot}$ black holes, and assume that accretion was triggered only by mergers. We find that these results agree well with our observed fraction.

There are some caveats to these results. We exclude ongoing mergers, where the AGN fraction may be higher. We did apply our X-ray analysis to the 43 galaxies that we excluded for being mergers. Of those, we found 2 to have $\geq 3\sigma$ X-ray sources: NMBS IDs 31041($\alpha=14^{\rm h} 17^{\rm m} 45.674^{\rm s}$, $\delta = +52^{\circ} 28^{\rm m} 2.29^{\rm s}$ (J2000)) \& 32018 ($\alpha=14^{\rm h} 18^{\rm m} 19.265^{\rm s}$, $\delta = +52^{\circ} 33^{\rm m} 51.464^{\rm s}$ (J2000)). This would increase our $\geq 3\sigma$ AGN fraction to a maximum of $\sim 2\%$. In addition, it is difficult to know the fraction of HMXBs in these galaxies as the effects of metallicity on the HMXB luminosity distribution are still unknown, and these are relatively metal poor systems compared to the galaxies typically studied. 

At this point, it is still too early to give any constraints on supermassive black hole seeds. However, it is clear that at least some $\sim10^9 \ M_{\odot}$ galaxies must have SMBHs. We hope to apply our technique to a larger cosmological volume and over a wider range of masses. In future work, we plan to explore how the predictions of semi-analytic models and numerical hydrodynamic simulations depend on black hole seeding mechanisms, as well as physical mechanisms for the triggering and regulation of black hole growth. This will allow us to properly compare the AGN fractions with other studies, and to give a constraint on seeding mechanisms.

\acknowledgements
The authors would like to thank the anonymous referee for their very thorough comments. This work is supported by \textit{Chandra} Archival Grant ARX 15700206. K.P. acknowledges support from the National Science Foundation Graduate Research Fellowship Program. Any opinions, findings, and conclusions or recommendations expressed in this material are those of the author(s) and do not necessarily reflect the views of the National Science Foundation. R.S.S. thanks the Downsbrough family for their generous support, and also gratefully acknowledges support from the Simons Foundation in the form of a Simons Investigator grant. R.C.H. acknowledges support from an Alfred P. Sloan Research Fellowship and a Dartmouth Class of 1962 Faculty Fellowship. Support for A.E.R. was provided by NASA through Hubble Fellowship grant HST-HF2-51347.001-A awarded by the Space Telescope Science Institute, which is operated by the Association of Universities for Research in Astronomy, Inc., for NASA, under contract NAS 5-26555. This research has made use of data obtained from the \textit{Chandra} Data Archive, the \textit{Chandra} Source Catalog, and software provided by the \textit{Chandra} X-ray Center (CXC) in the application packages CIAO, ChIPS, and Sherpa.

\clearpage

\begin{deluxetable*}{lccccccc}
\tablewidth{0pc}
\setlength{\tabcolsep}{.1mm}\tablecolumns{8}
\tablecaption{Basic Properties of $\geq 2\sigma$ Sources\label{tab-basicprop}}
\tablehead{
\colhead{NMBS ID} & \colhead{$\alpha_{\rm{J2000}}$\tablenotemark{a}} & \colhead{$\delta_{\rm{J2000}}$\tablenotemark{a}} & \colhead{$z_{spec}$\tablenotemark{b}} & \colhead{Stellar Mass\tablenotemark{c}} & \colhead{$m_K$\tablenotemark{d}} & \colhead{SFR\tablenotemark{e}} & \colhead{$L_{XB}$\tablenotemark{f}} \\
 \colhead{$\#$} & \colhead{deg} & \colhead{deg} & \colhead{} & \colhead{$\log(M/M_{\odot})$} & \colhead{mag} & \colhead{$\log(M_{\odot} \ \rm{yr}^{-1})$} & \colhead{$\log(\rm{erg \ s}^{-1})$}  }
\startdata
1789 & 214.24665 & 52.39860 & 0.2468 & 9.39 & 21.2 & -1.8 & 37.43\\
6614 & 214.40269 & 52.49404 & 0.8105 & 9.48 & 23.0 & 0.28 & 39.49\\
7013 & 214.27524 & 52.50460 & 0.8362 & 9.27 & 23.1 & -0.33 & 38.88\\
7290* & 214.37594 & 52.50924 & 0.1729 & 9.08 & 21.3 & -0.40 & 38.81\\
12516 & 214.48813 & 52.61451 & 0.7601 & 8.87 & 23.2 & -0.73 & 38.48\\
12753 & 214.61006 & 52.61919 & 0.8535 & 9.41 & 23.2 & -0.060 & 39.15\\
13373 & 214.69795 & 52.63038 & 0.3534 & 8.32 & 23.9 & -2.9 & 36.35\\
14021* & 214.65990 & 52.64013 & 0.2083 & 9.30 & 20.7 & -1.9 & 37.33\\
14708 & 214.59223 & 52.65499 & 0.5818 & 8.78 & 23.7 & -1.2 & 38.02\\
16867 & 214.79042 & 52.69311 & 0.4361 & 8.95 & 23.1 & -0.34 & 38.87\\
16879 & 214.65076 & 52.69407 & 0.4826 & 9.27 & 22.8 & -0.39 & 38.82\\
16986 & 214.54985 & 52.69630 & 0.8427 & 9.28 & 23.3 & -0.32 & 38.89\\
17650* & 214.51356 & 52.70632 & 0.1789 & 9.39 & 19.7 & 0.47 & 39.68\\
18139 & 214.73561 & 52.71528 & 0.7351 & 9.39 & 23.0 & -1.8 & 37.42\\
19869 & 214.71357 & 52.74905 & 0.4409 & 9.26 & 22.6 & -0.22 & 38.99\\
19957* & 214.76793 & 52.75079 & 0.3545 & 8.95 & 22.7 & -0.35 & 38.86\\
20697 & 214.86649 & 52.76676 & 0.4355 & 8.62 & 24.0 & -2.5 & 36.66\\
20960 & 214.88444 & 52.77121 & 0.4353 & 8.78 & 23.6 & -0.23 & 38.98\\
20987 & 214.76189 & 52.77085 & 0.7380 & 9.42 & 22.7 & 0.010 & 39.22\\
21055 & 214.82581 & 52.77209 & 0.2626 & 8.89 & 22.0 & -0.40 & 38.81\\
21091 & 214.78670 & 52.77206 & 0.2650 & 9.19 & 21.8 & -0.10 & 39.11\\
21469 & 214.86046 & 52.77822 & 0.3457 & 9.00 & 22.4 & -0.11 & 39.10\\
21690 & 214.86846 & 52.78437 & 0.7764 & 9.03 & 22.7 & 0.21 & 39.42\\
22579 & 214.90512 & 52.79933 & 0.1649 & 9.22 & 20.6 & -2.0 & 37.25\\
24295 & 214.81801 & 52.82905 & 0.7521 & 9.38 & 22.7 & -0.80 & 38.41\\
24680 & 214.85287 & 52.83023 & 0.5774 & 8.34 & 23.7 & 0.30 & 39.51\\
30536 & 214.42972 & 52.41506 & 0.7191 & 9.17 & 23.5 & 0.25 & 39.46\\
31018 & 214.49324 & 52.46432 & 0.2660 & 9.35 & 21.5 & -0.30 & 38.91\\
31097*$\dagger$ & 214.50167 & 52.47253 & 0.5321 & 8.74 & 24.1 & -0.73 & 38.48\\
31892 & 214.33956 & 52.55662 & 0.5722 & 9.41 & 22.8 & 0.22 & 39.43\\
32364*$\dagger$ & 214.52787 & 52.59865 & 0.2036 & 9.42 & 21.4 & -0.42 & 38.79\\
32682 & 214.46519 & 52.63130 & 0.1733 & 8.66 & 21.9 & -0.90 & 38.31\\
32708 & 214.40370 & 52.63011 & 0.8414 & 8.91 & 24.1 & 0.18 & 39.39\\
32820 & 214.52603 & 52.64448 & 0.2036 & 9.36 & 21.7 & -0.94 & 38.27\\
33817* & 214.79648 & 52.73136 & 0.08120 & 8.96 & 20.8 & -0.60 & 38.61\\
33865 & 214.74270 & 52.73832 & 0.3836 & 9.48 & 21.3 & -1.1 & 38.14\\
33915* & 214.53242 & 52.73998 & 0.06730 & 9.19 & 20.3 & -1.2 & 38.01\\
34168*$\dagger$ & 214.82505 & 52.76636 & 0.3457 & 8.53 & 23.1 & -1.0 & 38.18\\
34347* & 214.82049 & 52.78238 & 0.4544 & 9.48 & 21.9 & 0.11 & 39.32
\enddata

\tablenotetext{a}{Galaxy positions, as given by the NMBS}
\tablenotetext{b}{Spectroscopic Redshifts, as given by DEEP2}
\tablenotetext{c}{Stellar Mass, as given by the NMBS}
\tablenotetext{d}{Apparent K magnitude, as given by the NMBS}
\tablenotetext{e}{Star Formation Rate, as given by the NMBS}
\tablenotetext{f}{Estimated X-ray Luminosity from XBs, see Section 3.2 for a full description}
\tablenotetext{*}{$\geq 3\sigma$ sources}
\tablenotetext{$\dagger$}{Position reflects X-ray source position rather than optical position}
\end{deluxetable*}

\clearpage

\LongTables
\begin{landscape}
    \begin{deluxetable}{lccccccccccccc}
    \tablewidth{0pc}
    \setlength{\tabcolsep}{.1mm}\tablecolumns{14}
    \tablecaption{X-ray Properties of $\geq 2\sigma$ Sources\label{tab-xprop}}
    \tablehead{
    \colhead{ID} & \colhead{$t_{exp}$\tablenotemark{a}} & \colhead{$C_{0.5-2}$\tablenotemark{b}} & \colhead{$C_{2-7}$\tablenotemark{b}} & \colhead{$C_{0.5-7}$\tablenotemark{b}} & \colhead{$F_{0.5-2}$} & \colhead{$F_{2-7}$} & \colhead{$F_{0.5-7}$} & \colhead{HR\tablenotemark{c}} & \colhead{$P_{0.5-2}$\tablenotemark{d}} & \colhead{$P_{2-7}$\tablenotemark{d}} & \colhead{$P_{0.5-7}$\tablenotemark{d}} & \colhead{Sig Bands\tablenotemark{e}} & \colhead{$P_{\rm{XB}}$\tablenotemark{f}} \\
     \colhead{$\#$} & \colhead{ks} & \colhead{} & \colhead{} & \colhead{} & \colhead{$\rm{erg \ s}^{-1}  \rm{cm}^{-2}$} & \colhead{$\rm{erg \ s}^{-1} \rm{cm}^{-2}$} & \colhead{$\rm{erg \ s}^{-1}  \rm{cm}^{-2}$} & \colhead{} & \colhead{} & \colhead{} & \colhead{} & \colhead{} & \colhead{}  }
    \startdata
    1789 & 153 & $5_{\scriptscriptstyle-4}^{\scriptscriptstyle+3}$ & $<13$ & $9_{\scriptscriptstyle-6}^{\scriptscriptstyle+5}$ & $-15.7_{\scriptscriptstyle-0.7}^{\scriptscriptstyle+0.2}$ & $<-14.6$ & $-15.1_{\scriptscriptstyle-0.5}^{\scriptscriptstyle+0.2}$ & $-0.56_{\scriptscriptstyle-0.4}^{\scriptscriptstyle+0.3}$ & 0.0746 & 0.719 & 0.0416 & F  & $<1\times10^{-5}$\\
    6614 & 104 & $<14$ & $18_{\scriptscriptstyle-11}^{\scriptscriptstyle+10}$ & $20_{\scriptscriptstyle-13}^{\scriptscriptstyle+12}$ & $<-15.0$ & $-14.3_{\scriptscriptstyle-0.4}^{\scriptscriptstyle+0.2}$ & $-14.6_{\scriptscriptstyle-0.5}^{\scriptscriptstyle+0.2}$ & $0.88_{\scriptscriptstyle-0.2}^{\scriptscriptstyle+0.1}$ & 0.789 & 0.0340 & 0.0406 & H F  & $<1\times10^{-5}$\\
    7013 & 183 & $5_{\scriptscriptstyle-2}^{\scriptscriptstyle+2}$ & $1_{\scriptscriptstyle-3}^{\scriptscriptstyle+2}$ & $6_{\scriptscriptstyle-4}^{\scriptscriptstyle+3}$ & $-15.8_{\scriptscriptstyle-0.4}^{\scriptscriptstyle+0.1}$ & $-15.5^{\scriptscriptstyle+0.4}$ & $-15.4_{\scriptscriptstyle-0.4}^{\scriptscriptstyle+0.2}$ & $-0.53_{\scriptscriptstyle-0.5}^{\scriptscriptstyle+0.3}$ & 0.0104 & 0.304 & 0.0376 & S F  & $<1\times10^{-5}$\\
    7290* & 195 & $<0$ & $6_{\scriptscriptstyle-2}^{\scriptscriptstyle+2}$ & $4_{\scriptscriptstyle-2}^{\scriptscriptstyle+1}$ & $<-16.7$ & $-15.0_{\scriptscriptstyle-0.3}^{\scriptscriptstyle+0.1}$ & $-15.6_{\scriptscriptstyle-0.4}^{\scriptscriptstyle+0.1}$ & $0.96_{\scriptscriptstyle-0.07}^{\scriptscriptstyle+0.04}$ & 1.00 & $9.01\times 10^{-4}$ & $8.18\times 10^{-3}$ & H  & 0.0120\\
    12516 & 427 & $0_{\scriptscriptstyle-1}^{\scriptscriptstyle+0}$ & $6_{\scriptscriptstyle-3}^{\scriptscriptstyle+2}$ & $7_{\scriptscriptstyle-3}^{\scriptscriptstyle+3}$ & $-17.7^{\scriptscriptstyle+0.7}$ & $-15.4_{\scriptscriptstyle-0.4}^{\scriptscriptstyle+0.2}$ & $-15.6_{\scriptscriptstyle-0.3}^{\scriptscriptstyle+0.1}$ & $0.88_{\scriptscriptstyle-0.2}^{\scriptscriptstyle+0.1}$ & 0.586 & 0.0249 & $8.41\times 10^{-3}$ & H F  & $<1\times10^{-5}$\\
    12753 & 418 & $<0$ & $7_{\scriptscriptstyle-4}^{\scriptscriptstyle+3}$ & $9_{\scriptscriptstyle-4}^{\scriptscriptstyle+3}$ & $<-16.9$ & $-15.3_{\scriptscriptstyle-0.4}^{\scriptscriptstyle+0.2}$ & $-15.5_{\scriptscriptstyle-0.3}^{\scriptscriptstyle+0.1}$ & $0.94_{\scriptscriptstyle-0.1}^{\scriptscriptstyle+0.06}$ & 0.918 & 0.0334 & 0.0111 & H F  & $<1\times10^{-5}$\\
    13373 & 410 & $10_{\scriptscriptstyle-5}^{\scriptscriptstyle+4}$ & $<19$ & $2_{\scriptscriptstyle-9}^{\scriptscriptstyle+8}$ & $-15.7_{\scriptscriptstyle-0.4}^{\scriptscriptstyle+0.2}$ & $<-14.9$ & $-16.1^{\scriptscriptstyle+0.7}$ & $-0.82_{\scriptscriptstyle-0.2}^{\scriptscriptstyle+0.2}$ & 0.0223 & 0.925 & 0.423 & S  & $<1\times10^{-5}$\\
    14021* & 421 & $10_{\scriptscriptstyle-4}^{\scriptscriptstyle+3}$ & $6_{\scriptscriptstyle-5}^{\scriptscriptstyle+4}$ & $21_{\scriptscriptstyle-7}^{\scriptscriptstyle+6}$ & $-15.7_{\scriptscriptstyle-0.2}^{\scriptscriptstyle+0.1}$ & $-15.4_{\scriptscriptstyle-0.9}^{\scriptscriptstyle+0.2}$ & $-15.2_{\scriptscriptstyle-0.2}^{\scriptscriptstyle+0.1}$ & $-0.38_{\scriptscriptstyle-0.2}^{\scriptscriptstyle+0.4}$ & $3.53\times 10^{-4}$ & 0.117 & $5.58\times 10^{-4}$ & S F  & $<1\times10^{-5}$\\
    14708 & 352 & $2_{\scriptscriptstyle-2}^{\scriptscriptstyle+1}$ & $4_{\scriptscriptstyle-3}^{\scriptscriptstyle+2}$ & $7_{\scriptscriptstyle-4}^{\scriptscriptstyle+3}$ & $-16.3_{\scriptscriptstyle-0.8}^{\scriptscriptstyle+0.2}$ & $-15.4_{\scriptscriptstyle-0.7}^{\scriptscriptstyle+0.2}$ & $-15.5_{\scriptscriptstyle-0.4}^{\scriptscriptstyle+0.2}$ & $0.25_{\scriptscriptstyle-0.2}^{\scriptscriptstyle+0.6}$ & 0.0750 & 0.0900 & 0.0279 & F  & $<1\times10^{-5}$\\
    16867 & 568 & $<22$ & $28_{\scriptscriptstyle-16}^{\scriptscriptstyle+15}$ & $28_{\scriptscriptstyle-18}^{\scriptscriptstyle+17}$ & $<-15.5$ & $-14.8_{\scriptscriptstyle-0.4}^{\scriptscriptstyle+0.2}$ & $-15.2_{\scriptscriptstyle-0.5}^{\scriptscriptstyle+0.2}$ & $0.91_{\scriptscriptstyle-0.1}^{\scriptscriptstyle+0.09}$ & 0.835 & 0.0233 & 0.0408 & H F  & $<1\times10^{-5}$\\
    16879 & 399 & $10_{\scriptscriptstyle-5}^{\scriptscriptstyle+4}$ & $6_{\scriptscriptstyle-8}^{\scriptscriptstyle+7}$ & $14_{\scriptscriptstyle-10}^{\scriptscriptstyle+9}$ & $-15.7_{\scriptscriptstyle-0.3}^{\scriptscriptstyle+0.2}$ & $-15.3^{\scriptscriptstyle+0.3}$ & $-15.3_{\scriptscriptstyle-0.5}^{\scriptscriptstyle+0.2}$ & $-0.30_{\scriptscriptstyle-0.2}^{\scriptscriptstyle+0.4}$ & 0.0190 & 0.227 & 0.0691 & S  & $<1\times10^{-5}$\\
    16986 & 428 & $<4$ & $10_{\scriptscriptstyle-6}^{\scriptscriptstyle+5}$ & $6_{\scriptscriptstyle-6}^{\scriptscriptstyle+5}$ & $<-16.1$ & $-15.2_{\scriptscriptstyle-0.4}^{\scriptscriptstyle+0.2}$ & $-15.7_{\scriptscriptstyle-1}^{\scriptscriptstyle+0.3}$ & $0.87_{\scriptscriptstyle-0.2}^{\scriptscriptstyle+0.1}$ & 0.636 & 0.0311 & 0.141 & H  & $<1\times10^{-5}$\\
    17650* & 409 & $19_{\scriptscriptstyle-5}^{\scriptscriptstyle+4}$ & $5_{\scriptscriptstyle-5}^{\scriptscriptstyle+4}$ & $21_{\scriptscriptstyle-7}^{\scriptscriptstyle+6}$ & $-15.4_{\scriptscriptstyle-0.1}^{\scriptscriptstyle+0.09}$ & $-15.4_{\scriptscriptstyle-1}^{\scriptscriptstyle+0.3}$ & $-15.1_{\scriptscriptstyle-0.2}^{\scriptscriptstyle+0.1}$ & $-0.68_{\scriptscriptstyle-0.1}^{\scriptscriptstyle+0.2}$ & $<1\times10^{-5}$ & 0.140 & $8.20\times 10^{-4}$ & S F  & 0.124\\
    18139 & 588 & $0_{\scriptscriptstyle-8}^{\scriptscriptstyle+7}$ & $25_{\scriptscriptstyle-15}^{\scriptscriptstyle+14}$ & $18_{\scriptscriptstyle-17}^{\scriptscriptstyle+16}$ & $-16.9^{\scriptscriptstyle+0.9}$ & $-14.9_{\scriptscriptstyle-0.4}^{\scriptscriptstyle+0.2}$ & $-15.4_{\scriptscriptstyle-1}^{\scriptscriptstyle+0.3}$ & $0.80_{\scriptscriptstyle-0.2}^{\scriptscriptstyle+0.2}$ & 0.467 & 0.0277 & 0.112 & H  & $<1\times10^{-5}$\\
    19869 & 540 & $<13$ & $24_{\scriptscriptstyle-11}^{\scriptscriptstyle+10}$ & $16_{\scriptscriptstyle-12}^{\scriptscriptstyle+11}$ & $<-15.7$ & $-14.9_{\scriptscriptstyle-0.3}^{\scriptscriptstyle+0.1}$ & $-15.4_{\scriptscriptstyle-0.6}^{\scriptscriptstyle+0.2}$ & $0.92_{\scriptscriptstyle-0.1}^{\scriptscriptstyle+0.08}$ & 0.690 & $5.51\times 10^{-3}$ & 0.0754 & H  & $<1\times10^{-5}$\\
    19957* & 688 & $2_{\scriptscriptstyle-4}^{\scriptscriptstyle+3}$ & $29_{\scriptscriptstyle-9}^{\scriptscriptstyle+8}$ & $29_{\scriptscriptstyle-11}^{\scriptscriptstyle+10}$ & $-16.6^{\scriptscriptstyle+0.4}$ & $-14.9_{\scriptscriptstyle-0.2}^{\scriptscriptstyle+0.1}$ & $-15.2_{\scriptscriptstyle-0.2}^{\scriptscriptstyle+0.1}$ & $0.90_{\scriptscriptstyle-0.1}^{\scriptscriptstyle+0.1}$ & 0.306 & $7.50\times 10^{-5}$ & $1.27\times 10^{-3}$ & H F  & $<1\times10^{-5}$\\
    20697 & 639 & $<4$ & $13_{\scriptscriptstyle-6}^{\scriptscriptstyle+5}$ & $13_{\scriptscriptstyle-6}^{\scriptscriptstyle+5}$ & $<-16.2$ & $-15.2_{\scriptscriptstyle-0.3}^{\scriptscriptstyle+0.1}$ & $-15.5_{\scriptscriptstyle-0.3}^{\scriptscriptstyle+0.2}$ & $0.92_{\scriptscriptstyle-0.1}^{\scriptscriptstyle+0.08}$ & 0.659 & $6.09\times 10^{-3}$ & 0.0128 & H F  & $<1\times10^{-5}$\\
    20960 & 623 & $7_{\scriptscriptstyle-3}^{\scriptscriptstyle+2}$ & $3_{\scriptscriptstyle-5}^{\scriptscriptstyle+4}$ & $12_{\scriptscriptstyle-6}^{\scriptscriptstyle+5}$ & $-16.1_{\scriptscriptstyle-0.3}^{\scriptscriptstyle+0.1}$ & $-15.8^{\scriptscriptstyle+0.4}$ & $-15.5_{\scriptscriptstyle-0.3}^{\scriptscriptstyle+0.2}$ & $-0.45_{\scriptscriptstyle-0.08}^{\scriptscriptstyle+0.3}$ & $8.64\times 10^{-3}$ & 0.265 & 0.0144 & S F  & $2.66\times 10^{-5}$\\
    20987 & 676 & $2_{\scriptscriptstyle-3}^{\scriptscriptstyle+2}$ & $11_{\scriptscriptstyle-7}^{\scriptscriptstyle+6}$ & $8_{\scriptscriptstyle-7}^{\scriptscriptstyle+6}$ & $-16.6^{\scriptscriptstyle+0.3}$ & $-15.3_{\scriptscriptstyle-0.4}^{\scriptscriptstyle+0.2}$ & $-15.8_{\scriptscriptstyle-1}^{\scriptscriptstyle+0.3}$ & $0.68_{\scriptscriptstyle-0.2}^{\scriptscriptstyle+0.3}$ & 0.280 & 0.0340 & 0.124 & H  & $1.10\times 10^{-5}$\\
    21055 & 675 & $6_{\scriptscriptstyle-3}^{\scriptscriptstyle+3}$ & $<11$ & $8_{\scriptscriptstyle-5}^{\scriptscriptstyle+4}$ & $-16.1_{\scriptscriptstyle-0.4}^{\scriptscriptstyle+0.2}$ & $<-15.3$ & $-15.8_{\scriptscriptstyle-0.5}^{\scriptscriptstyle+0.2}$ & $-0.70_{\scriptscriptstyle-0.3}^{\scriptscriptstyle+0.3}$ & 0.0184 & 0.630 & 0.0688 & S  & $3.72\times 10^{-3}$\\
    21091 & 673 & $6_{\scriptscriptstyle-3}^{\scriptscriptstyle+3}$ & $<9$ & $6_{\scriptscriptstyle-6}^{\scriptscriptstyle+6}$ & $-16.2_{\scriptscriptstyle-0.5}^{\scriptscriptstyle+0.2}$ & $<-15.4$ & $-15.9^{\scriptscriptstyle+0.3}$ & $-0.72_{\scriptscriptstyle-0.3}^{\scriptscriptstyle+0.3}$ & 0.0409 & 0.751 & 0.168 & S  & 0.0418\\
    21469 & 628 & $5_{\scriptscriptstyle-3}^{\scriptscriptstyle+2}$ & $<8$ & $4_{\scriptscriptstyle-5}^{\scriptscriptstyle+4}$ & $-16.2_{\scriptscriptstyle-0.4}^{\scriptscriptstyle+0.2}$ & $<-15.4$ & $-16.0^{\scriptscriptstyle+0.3}$ & $-0.77_{\scriptscriptstyle-0.2}^{\scriptscriptstyle+0.3}$ & 0.0192 & 0.807 & 0.181 & S  & 0.0170\\
    21690 & 710 & $5_{\scriptscriptstyle-2}^{\scriptscriptstyle+2}$ & $6_{\scriptscriptstyle-4}^{\scriptscriptstyle+3}$ & $11_{\scriptscriptstyle-5}^{\scriptscriptstyle+4}$ & $-16.3_{\scriptscriptstyle-0.3}^{\scriptscriptstyle+0.1}$ & $-15.6_{\scriptscriptstyle-0.5}^{\scriptscriptstyle+0.2}$ & $-15.7_{\scriptscriptstyle-0.3}^{\scriptscriptstyle+0.1}$ & $0.012_{\scriptscriptstyle-0.4}^{\scriptscriptstyle+0.5}$ & $6.33\times 10^{-3}$ & 0.0479 & $9.88\times 10^{-3}$ & S F  & $1.59\times 10^{-5}$\\
    22579 & 682 & $5_{\scriptscriptstyle-3}^{\scriptscriptstyle+2}$ & $<6$ & $3_{\scriptscriptstyle-4}^{\scriptscriptstyle+3}$ & $-16.2_{\scriptscriptstyle-0.4}^{\scriptscriptstyle+0.1}$ & $<-15.6$ & $-16.2^{\scriptscriptstyle+0.3}$ & $-0.82_{\scriptscriptstyle-0.2}^{\scriptscriptstyle+0.2}$ & 0.0136 & 0.865 & 0.249 & S  & $<1\times10^{-5}$\\
    24295 & 713 & $3_{\scriptscriptstyle-2}^{\scriptscriptstyle+1}$ & $0_{\scriptscriptstyle-2}^{\scriptscriptstyle+1}$ & $3_{\scriptscriptstyle-3}^{\scriptscriptstyle+2}$ & $-16.5_{\scriptscriptstyle-0.4}^{\scriptscriptstyle+0.1}$ & $-17.2^{\scriptscriptstyle+1}$ & $-16.1_{\scriptscriptstyle-0.8}^{\scriptscriptstyle+0.2}$ & $-0.60_{\scriptscriptstyle-0.4}^{\scriptscriptstyle+0.3}$ & $9.09\times 10^{-3}$ & 0.526 & 0.104 & S  & $<1\times10^{-5}$\\
    24680 & 629 & $3_{\scriptscriptstyle-2}^{\scriptscriptstyle+1}$ & $<2$ & $4_{\scriptscriptstyle-3}^{\scriptscriptstyle+2}$ & $-16.4_{\scriptscriptstyle-0.4}^{\scriptscriptstyle+0.1}$ & $<-16.0$ & $-16.1_{\scriptscriptstyle-0.7}^{\scriptscriptstyle+0.2}$ & $-0.77_{\scriptscriptstyle-0.2}^{\scriptscriptstyle+0.3}$ & $7.97\times 10^{-3}$ & 0.703 & 0.0790 & S  & 0.0124\\
    30536 & 191 & $1_{\scriptscriptstyle-1}^{\scriptscriptstyle+0}$ & $1_{\scriptscriptstyle-1}^{\scriptscriptstyle+1}$ & $4_{\scriptscriptstyle-2}^{\scriptscriptstyle+1}$ & $-16.4^{\scriptscriptstyle+0.2}$ & $-15.7^{\scriptscriptstyle+0.2}$ & $-15.5_{\scriptscriptstyle-0.4}^{\scriptscriptstyle+0.1}$ & $-0.048_{\scriptscriptstyle-0.4}^{\scriptscriptstyle+0.6}$ & 0.123 & 0.225 & $9.12\times 10^{-3}$ & F  & $1.76\times 10^{-5}$\\
    31018 & 161 & $14_{\scriptscriptstyle-8}^{\scriptscriptstyle+7}$ & $7_{\scriptscriptstyle-14}^{\scriptscriptstyle+13}$ & $25_{\scriptscriptstyle-17}^{\scriptscriptstyle+16}$ & $-15.2_{\scriptscriptstyle-0.4}^{\scriptscriptstyle+0.2}$ & $-14.9^{\scriptscriptstyle+0.5}$ & $-14.7_{\scriptscriptstyle-0.5}^{\scriptscriptstyle+0.2}$ & $-0.30_{\scriptscriptstyle-0.2}^{\scriptscriptstyle+0.4}$ & 0.0292 & 0.296 & 0.0468 & S  & $<1\times10^{-5}$\\
    31097* & 66.3 & $3_{\scriptscriptstyle-2}^{\scriptscriptstyle+1}$ & $7_{\scriptscriptstyle-3}^{\scriptscriptstyle+2}$ & $10_{\scriptscriptstyle-3}^{\scriptscriptstyle+2}$ & $-15.5_{\scriptscriptstyle-0.4}^{\scriptscriptstyle+0.1}$ & $-14.4_{\scriptscriptstyle-0.2}^{\scriptscriptstyle+0.1}$ & $-14.7_{\scriptscriptstyle-0.2}^{\scriptscriptstyle+0.1}$ & $0.38_{\scriptscriptstyle-0.3}^{\scriptscriptstyle+0.3}$ & $2.15\times 10^{-3}$ & $4.00\times 10^{-6}$ & $<1\times10^{-5}$ & S H F  & $<1\times10^{-5}$\\
    31892 & 176 & $0_{\scriptscriptstyle-1}^{\scriptscriptstyle+1}$ & $8_{\scriptscriptstyle-4}^{\scriptscriptstyle+3}$ & $3_{\scriptscriptstyle-4}^{\scriptscriptstyle+3}$ & $-17.6^{\scriptscriptstyle+1}$ & $-14.9_{\scriptscriptstyle-0.3}^{\scriptscriptstyle+0.1}$ & $-15.6^{\scriptscriptstyle+0.3}$ & $0.87_{\scriptscriptstyle-0.2}^{\scriptscriptstyle+0.1}$ & 0.562 & $9.78\times 10^{-3}$ & 0.168 & H  & $2.18\times 10^{-4}$\\
    32364* & 425 & $9_{\scriptscriptstyle-3}^{\scriptscriptstyle+2}$ & $4_{\scriptscriptstyle-3}^{\scriptscriptstyle+2}$ & $10_{\scriptscriptstyle-4}^{\scriptscriptstyle+3}$ & $-15.8_{\scriptscriptstyle-0.2}^{\scriptscriptstyle+0.1}$ & $-15.5_{\scriptscriptstyle-0.5}^{\scriptscriptstyle+0.2}$ & $-15.5_{\scriptscriptstyle-0.2}^{\scriptscriptstyle+0.1}$ & $-0.39_{\scriptscriptstyle-0.3}^{\scriptscriptstyle+0.4}$ & $<1\times10^{-5}$ & 0.0420 & $1.91\times 10^{-4}$ & S F  & $1.68\times 10^{-3}$\\
    32682 & 379 & $2_{\scriptscriptstyle-2}^{\scriptscriptstyle+1}$ & $<4$ & $2_{\scriptscriptstyle-3}^{\scriptscriptstyle+2}$ & $-16.3_{\scriptscriptstyle-0.6}^{\scriptscriptstyle+0.2}$ & $<-15.4$ & $-16.1^{\scriptscriptstyle+0.3}$ & $-0.59_{\scriptscriptstyle-0.4}^{\scriptscriptstyle+0.3}$ & 0.0434 & 0.690 & 0.252 & S  & 0.0134\\
    32708 & 413 & $5_{\scriptscriptstyle-3}^{\scriptscriptstyle+2}$ & $1_{\scriptscriptstyle-5}^{\scriptscriptstyle+4}$ & $6_{\scriptscriptstyle-6}^{\scriptscriptstyle+5}$ & $-16.0_{\scriptscriptstyle-0.5}^{\scriptscriptstyle+0.2}$ & $-15.9^{\scriptscriptstyle+0.5}$ & $-15.6_{\scriptscriptstyle-1}^{\scriptscriptstyle+0.3}$ & $-0.40_{\scriptscriptstyle-0.6}^{\scriptscriptstyle+0.3}$ & 0.0430 & 0.388 & 0.132 & S  & $<1\times10^{-5}$\\
    32820 & 423 & $1_{\scriptscriptstyle-1}^{\scriptscriptstyle+0}$ & $1_{\scriptscriptstyle-3}^{\scriptscriptstyle+2}$ & $7_{\scriptscriptstyle-3}^{\scriptscriptstyle+3}$ & $-16.7^{\scriptscriptstyle+0.2}$ & $-16.1^{\scriptscriptstyle+0.4}$ & $-15.6_{\scriptscriptstyle-0.3}^{\scriptscriptstyle+0.2}$ & $0.12_{\scriptscriptstyle-1}^{\scriptscriptstyle+-0.9}$ & 0.198 & 0.372 & 0.0156 & F  & $9.64\times 10^{-5}$\\
    33817* & 684 & $14_{\scriptscriptstyle-6}^{\scriptscriptstyle+5}$ & $14_{\scriptscriptstyle-10}^{\scriptscriptstyle+9}$ & $30_{\scriptscriptstyle-13}^{\scriptscriptstyle+12}$ & $-15.8_{\scriptscriptstyle-0.2}^{\scriptscriptstyle+0.1}$ & $-15.2_{\scriptscriptstyle-0.5}^{\scriptscriptstyle+0.2}$ & $-15.2_{\scriptscriptstyle-0.2}^{\scriptscriptstyle+0.1}$ & $-0.13_{\scriptscriptstyle-0.3}^{\scriptscriptstyle+0.5}$ & $1.99\times 10^{-3}$ & 0.0594 & $3.92\times 10^{-3}$ & S  & 0.0318\\
    33865 & 627 & $13_{\scriptscriptstyle-7}^{\scriptscriptstyle+6}$ & $0_{\scriptscriptstyle-10}^{\scriptscriptstyle+9}$ & $7_{\scriptscriptstyle-12}^{\scriptscriptstyle+11}$ & $-15.8_{\scriptscriptstyle-0.3}^{\scriptscriptstyle+0.2}$ & $-16.5^{\scriptscriptstyle+1}$ & $-15.8^{\scriptscriptstyle+0.4}$ & $-0.59_{\scriptscriptstyle-0.4}^{\scriptscriptstyle+0.3}$ & 0.0159 & 0.485 & 0.272 & S  & $<1\times10^{-5}$\\
    33915* & 398 & $22_{\scriptscriptstyle-6}^{\scriptscriptstyle+5}$ & $2_{\scriptscriptstyle-8}^{\scriptscriptstyle+7}$ & $14_{\scriptscriptstyle-10}^{\scriptscriptstyle+9}$ & $-15.4_{\scriptscriptstyle-0.2}^{\scriptscriptstyle+0.1}$ & $-15.8^{\scriptscriptstyle+0.6}$ & $-15.3_{\scriptscriptstyle-0.5}^{\scriptscriptstyle+0.2}$ & $-0.79_{\scriptscriptstyle-0.2}^{\scriptscriptstyle+0.2}$ & $3.00\times 10^{-6}$ & 0.404 & 0.0649 & S  & $3.47\times 10^{-3}$\\
    34168* & 696 & $11_{\scriptscriptstyle-4}^{\scriptscriptstyle+3}$ & $15_{\scriptscriptstyle-6}^{\scriptscriptstyle+5}$ & $32_{\scriptscriptstyle-8}^{\scriptscriptstyle+7}$ & $-15.9_{\scriptscriptstyle-0.2}^{\scriptscriptstyle+0.1}$ & $-15.2_{\scriptscriptstyle-0.2}^{\scriptscriptstyle+0.1}$ & $-15.2_{\scriptscriptstyle-0.1}^{\scriptscriptstyle+0.09}$ & $0.11_{\scriptscriptstyle-0.2}^{\scriptscriptstyle+0.3}$ & $3.20\times 10^{-5}$ & $9.79\times 10^{-4}$ & $<1\times10^{-5}$ & S H F  & $<1\times10^{-5}$\\
    34347* & 602 & $19_{\scriptscriptstyle-4}^{\scriptscriptstyle+4}$ & $1_{\scriptscriptstyle-4}^{\scriptscriptstyle+3}$ & $19_{\scriptscriptstyle-6}^{\scriptscriptstyle+5}$ & $-15.6_{\scriptscriptstyle-0.1}^{\scriptscriptstyle+0.08}$ & $-16.1^{\scriptscriptstyle+0.5}$ & $-15.3_{\scriptscriptstyle-0.2}^{\scriptscriptstyle+0.1}$ & $-0.87_{\scriptscriptstyle-0.1}^{\scriptscriptstyle+0.1}$ & $<1\times10^{-5}$ & 0.360 & $6.80\times 10^{-5}$ & S F  & $6.77\times 10^{-5}$
    \enddata
    \tablenotetext{a}{Chandra Exposure Time at the Galaxy Position}
    \tablenotetext{b}{Net Source Counts in each band}
    \tablenotetext{c}{Hardness Ratios, as given by BEHR - see Section 2.2}
    \tablenotetext{d}{False Detection Probability for each band - see description in Section 2.2}
    \tablenotetext{e}{All Bands in which the source is significant}
    \tablenotetext{f}{Probability that the X-ray Luminosity observed could be due to an HXMB - see Section 3.2 for a full description}
    \tablenotetext{*}{$\geq 3\sigma$ sources}
    \end{deluxetable}
\clearpage
\end{landscape}

\begin{landscape}
    \begin{deluxetable}{lccccccccccc}
    \tablewidth{0pc}
    \setlength{\tabcolsep}{.1mm}\tablecolumns{12}
    \tablecaption{Basic Properties of $\geq 3\sigma$ Sources\label{tab-sigsources}}
    \tablehead{
    \colhead{ID} & \colhead{$\alpha_{\rm{J2000}}$} & \colhead{$\delta_{\rm{J2000}}$} & \colhead{$z_{spec}$} & \colhead{Stellar Mass} & \colhead{SFR} & \colhead{$L_{\rm{X}}$\tablenotemark{a}} & \colhead{HR\tablenotemark{b}} & \colhead{$P_{\rm{False}}$\tablenotemark{c}} & \colhead{$P_{\rm{XB}}$\tablenotemark{d}} & \colhead{[NII]/H$\alpha$\tablenotemark{e}} & \colhead{[OIII]/H$\beta$\tablenotemark{f}} \\
    \colhead{\#} & \colhead{deg} & \colhead{deg} & \colhead{} & \colhead{$\log(M/M_{\odot})$} &  \colhead{$\log(M_{\odot} \ \rm{yr}^{-1})$} & \colhead{$\log(\rm{erg \ s}^{-1})$} & \colhead{} & \colhead{} & \colhead{} & \colhead{} & \colhead{}}
    \startdata
    7290 & 214.37594 & 52.50924 & 0.1729 & 9.08 & -0.40 & $40.2_{\scriptscriptstyle-0.2}^{\scriptscriptstyle+0.2}$ & $0.96_{\scriptscriptstyle-0.07}^{\scriptscriptstyle+0.04}$ & $9.01\times 10^{-4}$ & 0.0120 & - & $4.31^{\star}$\\
    14021 & 214.65990 & 52.64013 & 0.2083 & 9.30 & -1.9 & $40.8_{\scriptscriptstyle-0.2}^{\scriptscriptstyle+0.1}$ & $-0.38_{\scriptscriptstyle-0.2}^{\scriptscriptstyle+0.4}$ & $5.58\times 10^{-4}$ & $<1\times10^{-5}$ & - & 2.34\\
    17650 & 214.51356 & 52.70632 & 0.1789 & 9.39 & 0.47 & $40.7_{\scriptscriptstyle-0.2}^{\scriptscriptstyle+0.1}$ & $-0.68_{\scriptscriptstyle-0.1}^{\scriptscriptstyle+0.2}$ & $8.20\times 10^{-4}$ & 0.124 & - & 3.36\\
    19957 & 214.76793 & 52.75079 & 0.3545 & 8.95 & -0.35 & $41.2_{\scriptscriptstyle-0.2}^{\scriptscriptstyle+0.1}$ & $0.90_{\scriptscriptstyle-0.1}^{\scriptscriptstyle+0.1}$ & $1.27\times 10^{-3}$ & $<1\times10^{-5}$ & - & 3.71\\
    31097$\dagger$ & 214.50167 & 52.47253 & 0.5321 & 8.74 & -0.73 & $42.1_{\scriptscriptstyle-0.1}^{\scriptscriptstyle+0.1}$ & $0.38_{\scriptscriptstyle-0.3}^{\scriptscriptstyle+0.3}$ & $<1\times10^{-5}$ & $<1\times10^{-5}$ & 0.16 & -\\
    32364$\dagger$ & 214.52787 & 52.59865 & 0.2036 & 9.42 & -0.42 & $40.5_{\scriptscriptstyle-0.2}^{\scriptscriptstyle+0.1}$ & $-0.39_{\scriptscriptstyle-0.3}^{\scriptscriptstyle+0.4}$ & $1.91\times 10^{-4}$ & $1.68\times 10^{-3}$ & 0.47 & -\\
    33817 & 214.79648 & 52.73136 & 0.08120 & 8.96 & -0.60 & $39.9_{\scriptscriptstyle-0.2}^{\scriptscriptstyle+0.2}$ & $-0.13_{\scriptscriptstyle-0.3}^{\scriptscriptstyle+0.5}$ & $1.99\times 10^{-3}$ & 0.0318 & 0.29 & -\\
    33915 & 214.53242 & 52.73998 & 0.06730 & 9.19 & -1.2 & $39.6_{\scriptscriptstyle-0.3}^{\scriptscriptstyle+0.3}$ & $-0.79_{\scriptscriptstyle-0.2}^{\scriptscriptstyle+0.2}$ & $<1\times10^{-5}$ & $3.47\times 10^{-3}$ & .21 & -\\
    34168$\dagger$ & 214.82505 & 52.76636 & 0.3457 & 8.53 & -1.0 & $41.2_{\scriptscriptstyle-0.1}^{\scriptscriptstyle+0.1}$ & $0.11_{\scriptscriptstyle-0.2}^{\scriptscriptstyle+0.3}$ & $<1\times10^{-5}$ & $<1\times10^{-5}$ & 0.19 & -\\
    34347 & 214.82049 & 52.78238 & 0.4544 & 9.48 & 0.11 & $41.4_{\scriptscriptstyle-0.1}^{\scriptscriptstyle+0.1}$ & $-0.87_{\scriptscriptstyle-0.1}^{\scriptscriptstyle+0.1}$ & $6.80\times 10^{-5}$ & $6.77\times 10^{-5}$ & 0.38 & -
    \enddata
    \tablenotetext{a}{X-ray Luminosity, as calculated using the net counts in the FB (0.5-7 keV)}
    \tablenotetext{b}{Hardness Ratios, as calculated using BEHR}
    \tablenotetext{c}{False Detection probability, see Section 2.2 for a full description}
    \tablenotetext{d}{Probability the source is a HMXB, see Section 3.2 for a full description}
    \tablenotetext{e}{[NII]/H$\alpha$ Line Ratio from the DEEP2 data}
    \tablenotetext{f}{[OIII]/H$\beta$ Line Ratio from the DEEP2 data}
    \tablenotetext{$\star$}{[OIII] from 4959\AA  line ([OIII]5007\AA = 3x[OIII]4959\AA)}
    \tablenotetext{$\dagger$}{Position reflects X-ray source position rather than optical position}
    \end{deluxetable}
\clearpage
\end{landscape}

\clearpage

\begin{deluxetable*}{lcccc}
\tablewidth{0pc}
\setlength{\tabcolsep}{.1mm}\tablecolumns{5}
\tablecaption{Observed AGN Fractions \label{tab-agnfrac}}
\tablehead{
\colhead{Study} & \colhead{Stellar Mass Range} & \colhead{Redshift Range} & \colhead{AGN Fraction} & \colhead{Parent Galaxy Selection Method}\\
 \colhead{} & \colhead{$\log(M/M_{\odot})$} & \colhead{} & \colhead{} & \colhead{}}
\startdata
This work & $<9.5$ & $0.1-0.6$ & $0.6-3\%$ & NMBS \\
Aird et al. (2012)\tablenotemark{$\dagger$} & $< 9.5$ & 0.2 - 0.6 & $1.1 - 4.5\%$ & PRIMUS \\
Miller et al. (2015) & $<10$ & $\lesssim 0.008$ & $>20\%$ & Virgo + Field\\
Schramm et al. (2013) & $<9.5$ & $<1$ & $\sim 0.1\%$ & GEMS\\
Lemons et al. (2015) & $<9.5$ & $<0.055$ & $\sim 2\%$ & NASA-Sloan
\enddata

\tablenotetext{$\dagger$}{Extrapolated from Aird et al. (2012), as discussed in Section 4.3}
\end{deluxetable*}


\begin{thebibliography}{}
\expandafter\ifx\csname natexlab\endcsname\relax\def\natexlab#1{#1}\fi

\bibitem[Abazajian et al.(2009)]{Abazajian2009} 
Abazajian, K.~N., Adelman-McCarthy, J.~K., Ag{\"u}eros, M.~A., et al.\ 2009, \apjs, 182, 543-558

\bibitem[{Aird {et~al.}(2012)Aird, Coil, Moustakas, Blanton, Burles, Cool,
  Eisenstein, Smith, Wong, \& Zhu}]{Aird2012}
Aird, J., Coil, A.~L., Moustakas, J., {et~al.} 2012, \apj, 746, 90

\bibitem[{Baldwin {et~al.}(1981)Baldwin, Phillips, \& Terlevich}]{Baldwin1981}
Baldwin, J.~A., Phillips, M.~M., \& Terlevich, R. 1981, \pasp, 93, 5

\bibitem[{Barth {et~al.}(2004)Barth, Greene, \& Ho}]{Barth2004}
Barth, A.~J., Greene, J.~E., \& Ho, L.~C. 2004, \apj, 619, 4

\bibitem[Basu-Zych et al.(2016)]{Basu2016}
Basu-Zych, A.~R., Lehmer, B., Fragos, T., et al.\ 2016, \apj, 818, 140 

\bibitem[Becerra et al.(2015)]{Becerra2015} 
Becerra, F., Greif, T.~H., Springel, V., \& Hernquist, L.~E.\ 2015, \mnras, 446, 2380 

\bibitem[{Bellovary {et~al.}(2011)Bellovary, Volonteri, Governato, Shen, Quinn,
  \& Wadsley}]{Bellovary2011}
Bellovary, J., Volonteri, M., Governato, F., {et~al.} 2011, \apj, 742,
  13

\bibitem[{Bertin \& Arnouts(1996)}]{Bertin1996}
Bertin, E., \& Arnouts, S. 1996, Astron. Astrophys. Suppl. Ser., 117, 393

\bibitem[{{Brandt} \& {Alexander}(2015)}]{Brandt2015}
{Brandt}, W.~N., \& {Alexander}, D.~M. 2015, \aapr, 23, 1

\bibitem[{{Brandt} \& {Hasinger}(2005)}]{Brandt2005}
{Brandt}, W.~N., \& {Hasinger}, G. 2005, \araa, 43, 827

\bibitem[{Bruzual \& Charlot(2003)}]{Bruzual2003}
Bruzual, G., \& Charlot, S. 2003, \mnras, 344, 1000

\bibitem[Chabrier(2003)]{Chabrier2003} 
Chabrier, G.\ 2003, \pasp, 115, 763 

\bibitem[{Davis {et~al.}(2002)Davis, Faber, Newman, Phillips, Ellis, Steidel,
  Conselice, Coil, Finkbeiner, Koo, Guhathakurta, Weiner, Schiavon, Willmer,
  Kaiser, Luppino, Wirth, Connolly, Eisenhardt, Cooper, \& Gerke}]{Davis2002}
Davis, M., Faber, S.~M., Newman, J.~a., {et~al.} 2002, Proc. SPIE, 4834, 12

\bibitem[{{Davis} {et~al.}(2007){Davis}, {Guhathakurta}, {Konidaris}, {Newman},
  {Ashby}, {Biggs}, {Barmby}, {Bundy}, {Chapman}, {Coil}, {Conselice},
  {Cooper}, {Croton}, {Eisenhardt}, {Ellis}, {Faber}, {Fang}, {Fazio},
  {Georgakakis}, {Gerke}, {Goss}, {Gwyn}, {Harker}, {Hopkins}, {Huang},
  {Ivison}, {Kassin}, {Kirby}, {Koekemoer}, {Koo}, {Laird}, {Le Floc'h}, {Lin},
  {Lotz}, {Marshall}, {Martin}, {Metevier}, {Moustakas}, {Nandra}, {Noeske},
  {Papovich}, {Phillips}, {Rich}, {Rieke}, {Rigopoulou}, {Salim},
  {Schiminovich}, {Simard}, {Smail}, {Small}, {Weiner}, {Willmer}, {Willner},
  {Wilson}, {Wright}, \& {Yan}}]{Davis2007}
{Davis}, M., {Guhathakurta}, P., {Konidaris}, N.~P., {et~al.} 2007, \apjl, 660,
  L1

\bibitem[{den Brok {et~al.}(2015)den Brok, Seth, Barth, {Carson, Daniel J.;
  Neumayer, Nadine; Cappellari, Michele; Debattista}, Ho, \& {Hood, Carol E.;
  McDermid}}]{denBrok2015}
den Brok, M., Seth, A.~C., Barth, A.~J., {et~al.} 2015, \apj, 801, 16

\bibitem[{Desroches \& Ho(2009)}]{Desroches2009}
Desroches, L.-B., \& Ho, L.~C. 2009, \apj, 690, 267

\bibitem[{Dong {et~al.}(2012)Dong, Ho, Yuan, Wang, Fan, Zhou, \&
  Jiang}]{Dong2012}
Dong, X.-B., Ho, L.~C., Yuan, W., {et~al.} 2012, \apj, 755, 167

\bibitem[Eisenstein \& Loeb(1995)]{Eisenstein1995} 
Eisenstein, D.~J., \& Loeb, A.\ 1995, \apj, 443, 11

\bibitem[{{Filippenko} \& {Sargent}(1989)}]{Filippenko1989}
{Filippenko}, A.~V., \& {Sargent}, W.~L.~W. 1989, \apjl, 342, L11

\bibitem[{{Fragos} {et~al.}(2013){Fragos}, {Lehmer}, {Tremmel}, {Tzanavaris},
  {Basu-Zych}, {Belczynski}, {Hornschemeier}, {Jenkins}, {Kalogera}, {Ptak}, \&
  {Zezas}}]{Fragos2013}
{Fragos}, T., {Lehmer}, B., {Tremmel}, M., {et~al.} 2013, \apj, 764, 41

\bibitem[{Fruscione {et~al.}(2006)Fruscione, McDowell, Allen, Brickhouse,
  Burke, \& Davis}]{Fruscione2006}
Fruscione, A., McDowell, J.~C., Allen, G.~E., {et~al.} 2006, Proc. SPIE, 6270

\bibitem[Gallo et al.(2008)]{Gallo2008} 
Gallo, E., Treu, T., Jacob, J., et al.\ 2008, \apj, 680, 154-168 

\bibitem[{Gebhardt {et~al.}(2001)Gebhardt, Lauer, Kormendy, Pinkney, Bower,
  Green, Gull, Hutchings, Kaiser, Nelson, Richstone, \&
  Weistrop}]{Gebhardt2001}
Gebhardt, K., Lauer, T.~R., Kormendy, J., {et~al.} 2001, \aj, 122, 2469

\bibitem[Gibson \& Brandt(2012)]{Gibson2012} 
Gibson, R.~R., \& Brandt, W.~N.\ 2012, \apj, 746, 54 

\bibitem[{Gilfanov {et~al.}(2004)Gilfanov, Grimm, \& Sunyaev}]{Gilfanov2004}
Gilfanov, M., Grimm, H.-J., \& Sunyaev, R. 2004, \mnras, 347, L57

\bibitem[Goswami et al.(2012)]{Goswami2012} 
Goswami, S., Umbreit, S., Bierbaum, M., \& Rasio, F.~A.\ 2012, \apj, 752, 43 

\bibitem[{{Goulding} \& {Alexander}(2009)}]{Goulding2009}
{Goulding}, A.~D., \& {Alexander}, D.~M. 2009, \mnras, 398, 1165

\bibitem[{Goulding {et~al.}(2012)Goulding, Forman, Hickox, Jones, Kraft,
  Murray, Vikhlinin, Coil, Cooper, Davis, \& Newman}]{Goulding2012}
Goulding, A.~D., Forman, W.~R., Hickox, R.~C., {et~al.} 2012, \apjs, 202, 26

\bibitem[Goulding et al.(2014)]{Goulding2014} 
Goulding, A.~D., Forman, W.~R., Hickox, R.~C., et al.\ 2014, \apj, 783, 40 

\bibitem[{Greene(2012)}]{Greene2012}
Greene, J.~E. 2012, Nat. Commun., 3, 1304

\bibitem[{Greene \& Ho(2004)}]{Greene2004}
Greene, J.~E., \& Ho, L.~C. 2004, \apj, 610, 13

\bibitem[{Greene \& Ho(2007)}]{Greene2007}
---. 2007, \apj, 13

\bibitem[{Greene {et~al.}(2010)Greene, Peng, Kuo, Braatz, Debattista, \&
  Popescu}]{Greene2010}
Greene, J.~E., Peng, C.~Y., Kuo, C.-Y., {et~al.} 2010, AIP Conf. Proc., 1240,
  207

\bibitem[{Groves {et~al.}(2006)Groves, Heckman, \& Kauffmann}]{Groves2006}
Groves, B.~a., Heckman, T.~M., \& Kauffmann, G. 2006, \mnras, 371, 1559

\bibitem[{{Hirschmann} {et~al.}(2012){Hirschmann}, {Naab}, {Somerville},
  {Burkert}, \& {Oser}}]{Hirschmann2012}
{Hirschmann}, M., {Naab}, T., {Somerville}, R.~S., {Burkert}, A., \& {Oser}, L.
  2012, \mnras, 419, 3200
  
\bibitem[Ho et al.(1997)]{Ho1997} 
Ho, L.~C., Filippenko, A.~V., \& Sargent, W.~L.~W.\ 1997, \apjs, 112, 315

\bibitem[{{Hopkins} {et~al.}(2007){Hopkins}, {Lidz}, {Hernquist}, {Coil},
  {Myers}, {Cox}, \& {Spergel}}]{Hopkins2007}
{Hopkins}, P.~F., {Lidz}, A., {Hernquist}, L., {et~al.} 2007, \apj, 662, 110

\bibitem[{{Hopkins} {et~al.}(2006){Hopkins}, {Somerville}, {Hernquist}, {Cox},
  {Robertson}, \& {Li}}]{Hopkins2006}
{Hopkins}, P.~F., {Somerville}, R.~S., {Hernquist}, L., {et~al.} 2006, \apj,
  652, 864
 
 \bibitem[Johnson et al.(2012)]{Johnson2012} 
 Johnson, J.~L., Whalen, D.~J., Fryer, C.~L., \& Li, H.\ 2012, \apj, 750, 66 

\bibitem[{Kalberla {et~al.}(2005)Kalberla, Burton, Hartmann, Arnal, Bajaja,
  Morras, \& P{\"{o}}ppel}]{Kalberla2005}
Kalberla, P. M.~W., Burton, W.~B., Hartmann, D., {et~al.} 2005, Astron.
  Astrophys., 440, 775
  
 \bibitem[Katz et al.(2015)]{Katz2015} 
 Katz, H., Sijacki, D., \& Haehnelt, M.~G.\ 2015, \mnras, 451, 2352 

\bibitem[{Kauffmann {et~al.}(2003{\natexlab{a}})Kauffmann, Heckman, Tremonti,
  Brinchmann, Charlot, White, Ridgway, Brinkmann, Fukugita, Hall, Ivezi{\'{c}},
  Richards, \& Schneider}]{Kauffmann2003}
Kauffmann, G., Heckman, T.~M., Tremonti, C., {et~al.} 2003{\natexlab{a}}, \mnras, 346, 1055

\bibitem[{Kauffmann {et~al.}(2003{\natexlab{b}})Kauffmann, Heckman, Tremonti,
  Brinchmann, Charlot, White, Ridgway, Brinkmann, Fukugita, Hall, Ivezi{\'{c}},
  Richards, \& Schneider}]{Kauffmann2003a}
---. 2003{\natexlab{b}}, \mnras, 346, 1055

\bibitem[{{Kewley} {et~al.}(2001){Kewley}, {Dopita}, {Sutherland}, {Heisler},
  \& {Trevena}}]{Kewley2001}
{Kewley}, L.~J., {Dopita}, M.~A., {Sutherland}, R.~S., {Heisler}, C.~A., \&
  {Trevena}, J. 2001, \apj, 556, 121

\bibitem[{{Kormendy} \& {Ho}(2013)}]{Kormendy2013}
{Kormendy}, J., \& {Ho}, L.~C. 2013, \araa, 51, 511

\bibitem[{Kriek {et~al.}(2009)Kriek, van Dokkum, Labbe, Franx, Illingworth,
  Marchesini, \& Quadri}]{Kriek2009}
Kriek, M., van Dokkum, P.~G., Labbe, I., {et~al.} 2009, \apj, 221

\bibitem[{{Laird} {et~al.}(2009){Laird}, {Nandra}, {Georgakakis}, {Aird},
  {Barmby}, {Conselice}, {Coil}, {Davis}, {Faber}, {Fazio}, {Guhathakurta},
  {Koo}, {Sarajedini}, \& {Willmer}}]{Laird2009}
{Laird}, E.~S., {Nandra}, K., {Georgakakis}, A., {et~al.} 2009, \apjs, 180, 102

\bibitem[Lang et al.(2014)]{Lang2014} 
Lang, D., Hogg, D.~W., \& Schlegel, D.~J.\ 2014, arXiv:1410.7397 

\bibitem[{Lehmer {et~al.}(2010)Lehmer, Alexander, Bauer, Brandt, Goulding,
  Jenkins, Ptak, \& Roberts}]{Lehmer2010}
Lehmer, B.~D., Alexander, D.~M., Bauer, F.~E., {et~al.} 2010, \apj,
  559

\bibitem[{Lemons {et~al.}(2015)Lemons, Reines, Plotkin, Gallo, \&
  Greene}]{Lemons2015}
Lemons, S., Reines, A., Plotkin, R., Gallo, E., \& Greene, J. 2015, \apj, 805, 10

\bibitem[Loeb \& Rasio(1994)]{Loeb1994}
Loeb, A., \& Rasio, F.~A.\ 1994, \apj, 432, 52

\bibitem[{{Ludwig} {et~al.}(2009){Ludwig}, {Wills}, {Greene}, \&
  {Robinson}}]{Ludwig2009}
{Ludwig}, R.~R., {Wills}, B., {Greene}, J.~E., \& {Robinson}, E.~L. 2009, \apj,
  706, 995
  
\bibitem[Madau \& Rees(2001)]{Madau2001} 
Madau, P., \& Rees, M.~J.\ 2001, \apjl, 551, L27 

\bibitem[{Marconi {et~al.}(2004)Marconi, Risaliti, Gilli, Hunt, Maiolino, \&
  Salvati}]{Marconi2004}
Marconi, A., Risaliti, G., Gilli, R., {et~al.} 2004, \mnras, 351, 169

\bibitem[Mayer et al.(2015)]{Mayer2015} 
Mayer, L., Fiacconi, D., Bonoli, S., et al.\ 2015, \apj, 810, 51 

\bibitem[{{McConnell} {et~al.}(2012){McConnell}, {Ma}, {Murphy}, {Gebhardt},
  {Lauer}, {Graham}, {Wright}, \& {Richstone}}]{McConnell2012}
{McConnell}, N.~J., {Ma}, C.-P., {Murphy}, J.~D., {et~al.} 2012, \apj, 756, 179

\bibitem[{{Mezcua} {et~al.}(2016){Mezcua}, {Civano}, {Fabbiano}, {Miyaji}, \&
  {Marchesi}}]{Mezcua2016}
{Mezcua}, M., {Civano}, F., {Fabbiano}, G., {Miyaji}, T., \& {Marchesi}, S.
  2016, \apj, 817, 20
 
 \bibitem[Miller et al.(2012)]{Miller2012} 
 Miller, B., Gallo, E., Treu, T., \& Woo, J.-H.\ 2012, \apj, 747, 57 

\bibitem[{Miller {et~al.}(2015)Miller, Gallo, Greene, Kelly, Treu, Woo, \&
  Baldassare}]{Miller2015}
Miller, B.~P., Gallo, E., Greene, J.~E., {et~al.} 2015, \apj, 799, 98

\bibitem[{Mineo {et~al.}(2012{\natexlab{a}})Mineo, Gilfanov, \&
  Sunyaev}]{Mineo2012}
Mineo, S., Gilfanov, M., \& Sunyaev, R. 2012{\natexlab{a}}, \mnras, 419, 2095

\bibitem[{Mineo {et~al.}(2012{\natexlab{b}})Mineo, Gilfanov, \&
  Sunyaev}]{Mineo2012b}
---. 2012{\natexlab{b}}, \mnras, 426, 1870

\bibitem[{{Moran} {et~al.}(2014){Moran}, {Shahinyan}, {Sugarman}, {V{\'e}lez},
  \& {Eracleous}}]{Moran2014}
{Moran}, E.~C., {Shahinyan}, K., {Sugarman}, H.~R., {V{\'e}lez}, D.~O., \&
  {Eracleous}, M. 2014, \aj, 148, 136

\bibitem[{{Moustakas} \& {Kennicutt}(2006)}]{Moustakas2006}
{Moustakas}, J., \& {Kennicutt}, Jr., R.~C. 2006, \apjs, 164, 81

\bibitem[{{Nandra} {et~al.}(2005){Nandra}, {Laird}, {Adelberger}, {Gardner},
  {Mushotzky}, {Rhodes}, {Steidel}, {Teplitz}, \& {Arnaud}}]{Nandra2005}
{Nandra}, K., {Laird}, E.~S., {Adelberger}, K., {et~al.} 2005, \mnras, 356, 568

\bibitem[{Nandra {et~al.}(2015)Nandra, Laird, Aird, Salvato, Georgakakis,
  Barro, Gonzalez, Barmby, Chary, Coil, Cooper, Davis, Dickinson, Faber, Fazio,
  Guhathakurta, Gwyn, Hsu, Huang, Ivison, Koo, Newman, Rangel, Yamada, \&
  Willmer}]{Nandra2015}
Nandra, K., Laird, E.~S., Aird, J.~a., {et~al.} 2015, \apjs, 220, 21

\bibitem[{{Newman} {et~al.}(2013){Newman}, {Cooper}, {Davis}, {Faber}, {Coil},
  {Guhathakurta}, {Koo}, {Phillips}, {Conroy}, {Dutton}, {Finkbeiner}, {Gerke},
  {Rosario}, {Weiner}, {Willmer}, {Yan}, {Harker}, {Kassin}, {Konidaris},
  {Lai}, {Madgwick}, {Noeske}, {Wirth}, {Connolly}, {Kaiser}, {Kirby},
  {Lemaux}, {Lin}, {Lotz}, {Luppino}, {Marinoni}, {Matthews}, {Metevier}, \&
  {Schiavon}}]{Newman2013}
{Newman}, J.~A., {Cooper}, M.~C., {Davis}, M., {et~al.} 2013, \apjs, 208, 5

\bibitem[Paggi et al.(2015)]{Paggi2015} 
Paggi, A., Fabbiano, G., Civano, F., et al.\ 2015, arXiv:1507.03170 

\bibitem[{Park {et~al.}(2006)Park, Kashyap, Siemiginowska, van Dyk, Zezas,
  Heinke, \& Wargelin}]{Park2006}
Park, T., Kashyap, V.~L., Siemiginowska, A., {et~al.} 2006, \apj, 652

\bibitem[Plotkin et al.(2016)]{Plotkin2016} 
Plotkin, R.~M., Gallo, E., Haardt, F., et al.\ 2016, arXiv:1605.00742 

\bibitem[Regan \& Haehnelt(2009)]{Regan2009} 
Regan, J.~A., \& Haehnelt, M.~G.\ 2009, \mnras, 393, 858 

\bibitem[{Reines \& Deller(2012)}]{Reines2012}
Reines, A.~E., \& Deller, A.~T. 2012, \apj, 750, L24

\bibitem[{Reines {et~al.}(2013)Reines, Greene, \& Geha}]{Reines2013}
Reines, A.~E., Greene, J.~E., \& Geha, M. 2013, \apj, 775, 116

\bibitem[{{Reines} {et~al.}(2014){Reines}, {Plotkin}, {Russell}, {Mezcua},
  {Condon}, {Sivakoff}, \& {Johnson}}]{Reines2014a}
{Reines}, A.~E., {Plotkin}, R.~M., {Russell}, T.~D., {et~al.} 2014, \apjl, 787, L30

\bibitem[{{Reines} {et~al.}(2011){Reines}, {Sivakoff}, {Johnson}, \&
  {Brogan}}]{Reines2011}
{Reines}, A.~E., {Sivakoff}, G.~R., {Johnson}, K.~E., \& {Brogan}, C.~L. 2011,
  \nat, 470, 66

\bibitem[{Reines \& Volonteri(2015)}]{Reines2015}
Reines, A.~E., \& Volonteri, M. 2015, \apj, 813, 82

\bibitem[Rix et al.(2004)]{Rix2004} 
Rix, H.-W., Barden, M., Beckwith, S.~V.~W., et al.\ 2004, \apjs, 152, 163

\bibitem[Sartori et al.(2015)]{Sartori2015} 
Sartori, L.~F., Schawinski, K., Treister, E., et al.\ 2015, \mnras, 454, 3722 

\bibitem[{Satyapal {et~al.}(2009)Satyapal, B{\"{o}}ker, Mcalpine, Gliozzi,
  Abel, \& Heckman}]{Satyapal2009}
Satyapal, S., B{\"{o}}ker, T., Mcalpine, W., {et~al.} 2009, \apj, 704, 439

\bibitem[{Satyapal {et~al.}(2008)Satyapal, Vega, Dudik, Abel, \&
  Heckman}]{Satyapal2008}
Satyapal, S., Vega, D., Dudik, R.~P., Abel, N.~P., \& Heckman, T. 2008, \apj, 677, 17

\bibitem[{Schramm {et~al.}(2013)Schramm, Silverman, Greene, Brandt, Luo, Xue,
  Capak, Kakazu, Kartaltepe, \& Mainieri}]{Schramm2013}
Schramm, M., Silverman, J.~D., Greene, J.~E., {et~al.} 2013, \apj, 773, 150

\bibitem[{{Seth} {et~al.}(2010){Seth}, {Cappellari}, {Neumayer}, {Caldwell},
  {Bastian}, {Olsen}, {Blum}, {Debattista}, {McDermid}, {Puzia}, \&
  {Stephens}}]{Seth2010}
{Seth}, A.~C., {Cappellari}, M., {Neumayer}, N., {et~al.} 2010, \apj, 714, 713

\bibitem[{Somerville {et~al.}(2008)Somerville, Hopkins, Cox, Robertson, \&
  Hernquist}]{Somerville2008}
Somerville, R.~S., Hopkins, P.~F., Cox, T.~J., Robertson, B.~E., \& Hernquist,
  L. 2008, \mnras, 391, 481
  
 \bibitem[Somerville et al.(2012)]{Somerville2012} 
 Somerville, R.~S., Gilmore, R.~C., Primack, J.~R., \& Dom{\'{\i}}nguez, A.\ 2012, \mnras, 423, 1992 

\bibitem[{Steidel {et~al.}(2003)Steidel, Adelberger, Shapley, Pettini,
  Dickinson, \& Giavalisco}]{Steidel2003}
Steidel, C.~C., Adelberger, K.~L., Shapley, A.~E., {et~al.} 2003, \apj, 592, 728

\bibitem[Stone et al.(2016)]{Stone2016} 
Stone, N.~C., Kuepper, A.~H.~W., \& Ostriker, J.~P.\ 2016, arXiv:1606.01909 

\bibitem[{Thornton {et~al.}(2008)Thornton, Barth, Ho, Rutledge, \&
  Greene}]{Thornton2008}
Thornton, C.~E., Barth, A.~J., Ho, L.~C., Rutledge, R.~E., \& Greene, J.~E.
  2008, \apj, 686, 892

\bibitem[{Valluri {et~al.}(2005)Valluri, Ferrarese, Merritt, \&
  Joseph}]{Valluri2005}
Valluri, M., Ferrarese, L., Merritt, D., \& Joseph, C.~L. 2005, \apj,
  628, 137

\bibitem[{{van den Bosch} \& {de Zeeuw}(2010)}]{van-den-Bosch2010}
{van den Bosch}, R.~C.~E., \& {de Zeeuw}, P.~T. 2010, \mnras, 401, 1770

\bibitem[{van Dokkum {et~al.}(2009)van Dokkum, Labb{\'{e}}, Marchesini,
  Quadri, Brammer, Whitaker, Kriek, Franx, Rudnick, Illingworth, Lee, \&
  Muzzin}]{vanDokkum2009}
van~van Dokkum, P.~G., Labb{\'{e}}, I., Marchesini, D., {et~al.} 2009, \pasp, 121, 2

\bibitem[{Volonteri(2010)}]{Volonteri2010}
Volonteri, M. 2010, \araa, 18, 279

\bibitem[{{Whalen} {et~al.}(2015){Whalen}, {Hickox}, {Reines}, {Greene},
  {Sivakoff}, {Johnson}, {Alexander}, \& {Goulding}}]{Whalen2015}
{Whalen}, T.~J., {Hickox}, R.~C., {Reines}, A.~E., {et~al.} 2015, \apj, 806, 37

\bibitem[{Whitaker {et~al.}(2011)Whitaker, Labbe, van Dokkum, Brammer, Kriek,
  Marchesini, Quadri, Franx, Muzzin, Williams, Bezanson, Illingworth, Lee,
  Lundgren, Nelson, Rudnick, Tal, \& Wake}]{Whitaker2011}
Whitaker, K.~E., Labbe, I., van Dokkum, P.~G., {et~al.} 2011, \apj, 86, 24

\bibitem[Windhorst et al.(2009)]{Windhorst2009} 
Windhorst, R.~A., Mather, J., Clampin, M., et al.\ 2009, astro2010: The Astronomy and Astrophysics Decadal Survey, 2010 

\bibitem[Xue et al.(2016)]{Xue2016} 
Xue, Y.~Q., Luo, B., Brandt, W.~N., et al.\ 2016, arXiv:1602.06299 

\end{thebibliography}
\end{document}